\documentclass{aa}
\usepackage{pdflscape}
\usepackage{soul}
\usepackage{color}
\usepackage{times}
\usepackage{natbib}
\usepackage{amsmath,amssymb,amsfonts,txfonts}
\usepackage{multirow}
\usepackage{cellspace, hhline}
\usepackage{nicefrac}
\usepackage{notes2bib}
\usepackage{graphicx}
\usepackage{float}
\usepackage{tabularx,ragged2e}
\newcolumntype{C}{>{\Centering\arraybackslash}X} 
\usepackage{txfonts}
\usepackage{multirow}
\usepackage{cellspace, hhline}
\usepackage{nicefrac}
\usepackage{notes2bib}
\usepackage{verbatim}
\usepackage{makecell}
\usepackage{bm}

\usepackage{hyperref} 
\hypersetup{colorlinks=true,linkcolor=[rgb]{1.,0.2,0.2},citecolor=[rgb]{0.1,0.4,1.},filecolor=[rgb]{0.7,0.2,0.2},urlcolor=[rgb]{0.7,0.2,0.2}}

\definecolor{darkspringgreen}{rgb}{0.09, 0.45, 0.27}
\newcommand{\adriano}[1]{{#1}} %Adriano
\newcommand{\nikki}[1]{{#1}} %Nikki
\newcommand{\rev}[1]{{#1}} % for the second round of revisions

\usepackage{etoolbox}

\begin{document}

   \title{Cosmic dissonance: new physics or \\ systematics behind a short sound horizon?} 
   \titlerunning{A Short Sound Horizon, new physics or systematics?}

   \author{Nikki Arendse\inst{\ref{dark}}\thanks{email: nikki.arendse@nbi.ku.dk ; }
          \and
          Rados{\l}aw ~J. ~Wojtak\inst{\ref{dark}}
          \and
          Adriano Agnello\inst{\ref{dark}}
          \and
          Geoff C.-F. Chen\inst{\ref{ucd}}
          \and 
          Christopher~D. Fassnacht\inst{\ref{ucd}}
          \and
          Dominique Sluse\inst{\ref{liege}}
          \and
          Stefan Hilbert\inst{\ref{ec},\ref{usm}}
          \and
          Martin Millon\inst{\ref{lausanne}}
          \and
          Vivien Bonvin\inst{\ref{lausanne}}
          \and
          Kenneth C. Wong$^7$
          \and
          Fr{\'e}d{\'e}ric Courbin$^6$
          \and 
          Sherry~H. Suyu$^{8,9,10}$
          \and
          Simon Birrer$^{11,12}$
          \and
          Tommaso Treu$^{11}$
          \and
          Leon~V.E. Koopmans$^{13}$
          }
   \authorrunning{Arendse, Wojtak, Agnello, et al.}

   \institute{
             DARK, Niels Bohr Institute, Lyngbyvej 2, 2100 Copenhagen, Denmark \label{dark}\and
             Physics Department UC Davis, 1 Shields Ave., Davis, CA 95616, USA \label{ucd} \and
STAR Institute, Quartier Agora - All\'ee du six Ao\^ut, 19c B-4000 Li\`ege, Belgium \label{liege}
\and
Exzellenzcluster Universe, Boltzmannstr. 2, D-85748 Garching, Germany \label{ec}\and
 Ludwig-Maximilians-Universit{\"a}t, Universit{\"a}ts-Sternwarte, Scheinerstr. 1, D-81679 M{\"u}nchen, Germany \label{usm}\and
 Laboratoire d'Astrophysique, {\'E}cole Politechnique F{\'e}d{\'e}rale de Lausanne (EPFL), Obs. de Sauverny, 1290 Versoix, Switzerland \label{lausanne}\\
             \noindent $^{7-12}$ the full list of affiliations can be found at the end of this paper
             }

\date{Received September 17, 2019; accepted May 7, 2020}

% \abstract{}{}{}{}{} 
% 5 {} token are mandatory

  \abstract
  % Context
  {Persistent tension between low-redshift observations and the Cosmic Microwave Background radiation (CMB), in terms of two fundamental distance scales set by the sound horizon ${r_{\rm d}}$ and the Hubble constant ${H_0}$, suggests new physics beyond the Standard Model, departures from concordance cosmology, or residual systematics.}
  % Aims
  {{The role of different probe combinations must be assessed, as well as of different physical models that can alter the expansion history of the Universe and the inferred cosmological parameters.}}
  % Methods
  {We examine recently updated distance calibrations from Cepheids, gravitational lensing time-delay observations, and the Tip of the Red Giant Branch. {Calibrating the Baryon Acoustic Oscillations (BAO) and Type Ia supernovae with combinations of the distance indicators, we obtain a joint and self-consistent measurement of $H_0$ and $r_{\rm d}$ at low redshift, independent of cosmological models and CMB inference. In an attempt to alleviate the tension between late-time and CMB-based measurements, we consider four extensions of the standard $\Lambda$CDM model.} }
  % Results
  {{The sound horizon from our different measurements is ${r_{\rm d}=(137\pm3^{stat.}\pm 2^{syst.})}$~Mpc based on absolute distance calibration from gravitational lensing and the cosmic distance ladder. Depending on the adopted distance indicators, the \textit{combined} tension in $H_0$ and $r_{\rm d}$ ranges between 2.3 and 5.1 $\sigma$, and is independent of changes to the low-redshift expansion history. We find that modifications of $\Lambda$CDM that change the physics after recombination fail to provide a solution to the problem, for the reason that they only resolve the tension in $H_0$, while the tension in $r_{\rm d}$ remains unchanged. Pre-recombination extensions (with early dark energy or the effective number of neutrinos ${\textrm{N}_{\textrm{eff}}=3.24 \pm 0.16}$)} are allowed by the data, unless the calibration from Cepheids is included.}
  % Conclusions
  {Results from time-delay lenses are consistent with those from distance-ladder calibrations {and point to a discrepancy between absolute distance scales measured from the CMB (assuming the standard cosmological model) and late-time observations.} New proposals to resolve this tension {should be examined with respect to reconciling not only the Hubble constant but also the sound horizon derived from the CMB and other cosmological probes.}}
  

   \maketitle

\section{Introduction}
At the onset of matter-radiation decoupling after the Big Bang, {photon-baryon fluid underwent oscillations whose characteristic physical scale is described by the so-called \textit{sound horizon} $r_{\rm s}$. This leaves a characteristic imprint on large scale distribution of baryons, with its characteristic size fixed in the comoving coordinates and equal to the sound horizon at the drag epoch $z_{\rm d}$ given by}
{
\begin{equation}
{r_{\rm d}\equiv r_{\rm s}(z_{\rm d})=\int_{z_{\rm d}}^{\infty}\frac{c_{\rm s}\mathrm{d}z}{H(z)}\ \ ,}
\end{equation}
where $c_{s}$ is the sound speed in the primordial plasma and $H(z)$ is the Hubble parameter.
}

The sound horizon $r_{\rm d}$ is robustly determined from the Cosmic Microwave Background measurements \nikki{(CMB)}, if the Standard Model of particle physics as well as the {standard cosmological model in the pre-recombination Universe} are adopted \citep{Planck2018}. Alternatively, it can be measured at later times, from the Baryon Acoustic Oscillation (BAO) peak in the two-point spatial correlation function of galaxies and quasars. The latter is an angular measurement, which can be converted into a physical $r_{\rm d}$ measurement through independent distance calibrations \citep[see e.g.][]{Hea2014,Ber2016,Ver2017,Arendse2019,Ayl2019}. \nikki{The parameter $r_{\rm d}$ is intimately linked to the current expansion rate of the Universe, the Hubble constant $H_0$, since BAO measurements constrain the product of $H_0$ and $r_{\rm d}$.}

Accurate {distance measurements from CMB-independent observations} 
can be used to determine $r_{\rm d}$ \nikki{and $H_0$} \nikki{in a way that is} truly independent of early-Universe physics. \nikki{Therefore, these measurements can} test our understanding of the concordance cosmology and the Standard  Model of particle physics\adriano{, through low-redshift measurements only}. {Type Ia supernovae, calibrated by} Cepheids {with three independent distance anchors (parallaxes in the Milky Way, detached eclipsing binaries in the LMC and maser galaxy NGC 4258),} provide the most precise distance calibration to date, as performed by the \textit{Supernovae and $H_{0}$ for the Equation of State of dark energy} project \cite[SH0ES; ][]{Riess2019}. Another powerful way of obtaining absolute distances is by using strongly lensed quasar systems, which extend to higher redshifts than the Cepheids. The \textit{$H_0$ Lenses in COSMOGRAIL’s Wellspring} collaboration 
{\cite[H0LiCOW, ][]{Suyu2017}} has provided \adriano{few-}percent-level precision constraints on $H_0$ from time-delay cosmology. \adriano{Over the whole sample, the effect of known systematics is at $\lesssim1\%$ level, currently negligible with respect to the statistical uncertainties \citep{Millon20}.}
The latest results from SH0ES and H0LiCOW indicate a strong tension in the Hubble constant $H_0$ between \nikki{late-time observations (CMB-independent probes including primarily type Ia SNe, lensing and BAO) and CMB-based measurements}, within a flat-$\Lambda$CDM model. Previous results based on four lenses alone \citep{Arendse2019, Taubenberger2019} resulted in a $2\sigma$ \adriano{discrepancy}, while a six-lens analysis \citep{Wong2019} gave a $3\sigma$ tension. 
When combined with the distance-ladder results by SH0ES, the tension increases to a $5\sigma$ level, still adopting a flat $\Lambda$CDM cosmological model. It is worth noting that the tension between the late-time and CMB-based measurements of $H_{0}$ is mildly lowered by the recent measurement making use of precise distance calibration from the Tip of the Red Giant Branch {(TRGB), as measured by the Carnegie-Chicago Hubble Project  \citep[herafter CCHP,][]{Freedman2019}}. These measurements fall between those from SH0ES and the CMB, at $1.7\sigma$ and $1.2\sigma$ differences respectively. \rev{For the sake of completeness, it is also worth mentioning that the Planck value of the Hubble constant is recovered in a CMB-independent but model-dependent analysis of BAO observations with the prior on the baryon density from the standard Big Bang Nucleosynthesis \citep{Cuceu2019,Addison2018}.}

\nikki{In this work, we revisit the claimed tension between late-time observations and the CMB in terms of the sound horizon and the Hubble constant, by making use of recent updated distance calibrations from gravitational time-delay lenses (H0LiCOW), Cepheids (SH0ES), and TRGB (CCHP). Through our methods (summarised in section \ref{sect:models}), we obtain measurements of $r_{\rm d}$ for different combinations of late-time distance calibrations in a manner that is almost completely independent of any cosmological model. 
Moreover, we investigate selected extensions to the standard $\Lambda$CDM model that have recently been proposed as possible solutions to the Hubble tension. Such new models attempt to reconcile the tension by modifying the expansion history of the standard model either before or after recombination, hereafter \textit{early-time} and \textit{late-time} modifications, and 
thus increasing the Hubble constant derived from the CMB.
We demonstrate that the late-time extensions fail to provide a solution to the problem, for the reason that they only succeed in alleviating the tension in $H_0$, while the tension in $r_{\rm d}$ remains unchanged. Our analysis emphasizes the importance of \rev{comparing at least $H_0$ and $r_{\rm d}$ derived from late-time observations and the CMB} when testing new models devised to mitigate the Hubble constant tension.
}

\nikki{This paper is structured as follows. Section \ref{sect:latetime} describes the late-time measurements of $r_{\rm d}$ and $H_0$, including the different data sets, models and inference methods that are used. In section \ref{sect:CMBcomp} \adriano{we outline how the \nikki{late-time} measurements are compared with CMB inference and extensions of the concordance scenario.} Our results are described in section \ref{sect:results} and our conclusions in section \ref{sect:conclusions}.}

{\section{Late-time measurements: data and methods}\label{sect:latetime}}

\noindent The values of $r_{\rm d}$ and $H_0$ can be constrained by employing several CMB-independent probes \adriano{at $0<z<2$}, in this paper referred to as late-time measurements. In section \ref{sect:datasets}, we provide an overview of the data sets \adriano{that} we use in our analysis. Section \ref{sect:models} introduces the models we choose to fit the Hubble diagram and interpolate up to redshift zero. By choosing models that are independent of cosmology we minimise the systematic uncertainty associated with cosmological model choices. 
Details about the inference are discussed in section \ref{sect:inference}\adriano{, and functional tests are shown in Appendix~\ref{Appendix:parametrizations}}. \\

\subsection{Data sets}\label{sect:datasets}

\noindent \nikki{The shape of the late-time expansion of the Universe \adriano{has been mapped precisely with} type Ia supernovae (SNe). In this work, we use relative distance moduli from the Pantheon sample \citep{Scol2018}.}

\nikki{Information about $r_{\rm d}$ is introduced by adding BAO measurements, which constrain the product of $H_0$ and $r_{\rm d}$. Our main results are obtained for the Hubble parameters $H(z)$ and the transverse comoving distances $D_{\rm M}(z)$ determined from the Baryon Oscillations Spectroscopic Survey \citep[BOSS;][]{Alam2017}. Additionally, we look into the effect of adding BAO constraints from the correlation of Ly$\alpha$ forest absorption and quasars in the extended Baryon Oscillation Spectroscopic Survey \citep[eBOSS;][]{Ag2019, Blom2019} and several isotropic BAO measurements. The isotropic measurements do not contain sufficient statistics to measure $H(z)$ and $D_{\rm M} (z)$ separately, but combine them in the volume-averaged distance $D_{\rm V} = \left( c z D_{\rm M}^2(z) H^{-1}(z) \right)^{1/3}$. We include two measurements from the reconstructed 6-degree Field Galaxy Survey  \citep{Carter2018}, two from eBOSS by \citep{Bautista2018,Ata2018}, and three from the WiggleZ Dark Energy Survey \citep{Kazin2019}. }

\nikki{Both SNe and BAO measurements provide only relative distances, thus their distance scale needs to be calibrated with absolute distance measurements.} Time-delay and angular-diameter distances to strongly lensed quasars, obtained by the H0LiCOW collaboration, {provide such an absolute calibration of cosmological distances} \citep[see e.g.][and references therein]{Suyu2017}. Results from a fifth and a sixth lensed quasar system have been recently obtained {\citep{cfc2019,Rusu2019,Bonvin2019,Sluse2019}, including new distance measurements on previous lensed quasar systems using new data and analysis \citep{cfc2019,Jee2019}.} {In this work, we use complete constraints on distances from observations of the 6 lensed quasars systems, as summarized in \cite{Wong2019}}. The information from the lensed quasars is modeled self-consistently, together with the relative distance indicators (SNe, BAO). \\

\noindent {Keeping the lensing data as our primary calibration of the absolute distance scale in all fits, we also include two optional priors given by recent local determinations of the Hubble constant. The first is the latest SH0ES measurement yielding $H_0 = $ 74.03  $\pm$ 1.42 km~s$^{-1}$~Mpc$^{-1}$ \citep{Riess2019}. The second is based on calibrating distances with the Tip of the Red Giant Branch (TRGB), a standard candle alternative to Cepheids. Here, analyses carried out by two separate groups have resulted in different values for $H_0$: 
\rev{\cite{Yuan2019} found 72.4 $\pm$ 2.0 km~s$^{-1}$~Mpc$^{-1}$, while CCHP obtained 69.6 $\pm$ 2.0 km~s$^{-1}$~Mpc$^{-1}$ \citep{Freedman2019,Freedman2020}.} In order to include both the highest and lowest late-time measurements of $H_0$, we have chosen to use the CCHP results for the TRGB and SH0ES results for Cepheids in our analysis. Since there is a partial overlap in the galaxy samples considered for the TRGB and Cepheid measurements, the two calibrations will only be applied separately.}

\noindent \nikki{Finally,} quasars { are optionally used as secondary} standard candles at high redshifts, by means of a relation between their UV and X-ray luminosities \citep{Ris2018}. {We do so in one of our inference runs in Table \ref{table3}, as an independent check.} \\

\noindent Our constraints on the late-time expansion is largely based on datasets and models that we explored in previous work \citep{Arendse2019}. The difference with previous datasets is the inclusion of two additional quasar-lens measurements {\citep{cfc2019,Rusu2019}}, Ly$\alpha$ BAO measurements at $z=2.34$ and $2.35$, several volume-averaged BAO measurements ($D_{\rm V}$ BAO), and the combination with the Cepheid distance-ladder or the TRGB calibration. \\

\subsection{\nikki{Models}}\label{sect:models}

\nikki{{Measuring $r_{\rm d}$ and $H_0$ from the observations described above requires adopting a model of the expansion history.} 
This is usually done by means of employing the standard $\Lambda$CDM model,} but any tension among different $r_{\rm d}$ and $H_0$ measurements in the $\Lambda$CDM framework may mean that the  $\Lambda$CDM expansion history is not necessarily an adequate model choice. \nikki{Instead of employing different extensions to $\Lambda$CDM to overcome this issue, we use three different models of polynomial parametrizations, that are completely agnostic about the underlying expansion history. This allows us to make an inference of $r_{\rm d}$ and $H_0$ that is based solely on observational data, and does not rely on cosmology.}

\nikki{The specifications of the three polynomial parametrizations (hereafter referred to as model 1, 2 and 3) are listed in Table~\ref{table:models}.
Model 1 adopts a polynomial expansion of $H(z)$ \citep{Weinberg1972,Visser2004}, model 2 expands the luminosity distance $D_{\rm{L}}$\footnote{\rev{Where the distance measures are related to each other according to $D_L = (1+z) D_M = (1+z)^2 D_A$.}} as a polynomial in $\log(1+z)$ \citep{Ris2018}, and model 3 describes transverse comoving distances $D_{\rm M}$ by polynomials in $z/(1+z)$ \citep{Cat2007,Li2019a}. For model 1, comoving distances are obtained from $H(z)$ through direct numerical integration of
\begin{equation}
d_{\rm c} (z)= \int_0^z \frac{c}{H(z)} \; \rm{d}z,
\end{equation}
and for model 2 and 3, $H(z)$ is obtained through
\begin{equation}
H(z, \Omega_k) = \frac{\mathrm{c}}{\partial D_M(z)/\partial z} \; \sqrt{1 + \frac{H_0^2 \Omega_k}{\mathrm{c}^2} D_{\rm M}(z)^2}\
\end{equation}
\noindent\citep{Weinberg1972}}.

We truncate all polynomials at the lowest expansion order required by the condition that model 1, 2 and 3 recover distances in a $\Lambda$CDM model, if their free coefficients are fixed at values found by Taylor expanding the corresponding functions in the fiducial $\Lambda$CDM model (see more in Appendix~\ref{Appendix:parametrizations}). This guarantees that expansion histories derived from the employed models converge to $\Lambda$CDM once observations become consistent exclusively with the standard model. Distances in $\Lambda$CDM are recovered with a minimum accuracy of 2 percent at $z<1.8$, where the accuracy limit is set by the current precision of the Hubble constant measurements and the upper limit of redshift is given by the most distant lensed quasar. Including higher order terms is disfavoured by Bayesian Information Criterion (BIC). \rev{In Appendix \ref{Appendix:parametrizations} we also show that this convergence criterion ensures that biases in $H_0$ are at sub-percent level, and biases in $q_0$ at a few-percent level.}

Finally, \nikki{in order to compare models 1-3 with the most commonly adopted cosmological model,} the fourth family {(model 4)} adopts a $\Lambda$CDM parameterisation. In all cases, both flatness and departures from it are considered.

\setlength{\tabcolsep}{0.5em} 
{\renewcommand{\arraystretch}{2.3}
\begin{table}[]
\begin{tabular}{l|l}
\nikki{\textbf{Model}}  & \nikki{\textbf{Formula}}   \\
\hline
\nikki{\textbf{1}} & \nikki{${ H(z) = H_0 \left( 1 + b_1 z + b_2 z^2 \right)}$} \\
\nikki{\textbf{2}} &  \makecell[l]{\nikki{${D_L (z) = \frac{c\ln(10)}{H_0} \left( \log(1+z) + c_2 [\log(1+z)]^2 \right.}$} \\ \nikki{\hspace{1.1cm} ${ \left. + \, c_3 [\log(1+z)]^3 + c_4 [\log(1+z)]^4 \right) }$}} \\
\nikki{\textbf{3}}   & \nikki{${D_M (z) = \frac{c}{H_0} \left( \frac{z}{1+z} + d_2 \left[\frac{z}{1+z}\right]^2 + d_3 \left[\frac{z}{1+z}\right]^3 + d_4 \left[\frac{z}{1+z}\right]^4 \right)}$} \\
\nikki{\textbf{4}}   & \nikki{${H(z) = H_0 \sqrt{\Omega_M \, (1+z)^3 + \Omega_{\Lambda} + \Omega_k \, (1+z)^2} } $} \\
\hline
\end{tabular}
\caption{
{\nikki{The three polynomial parametrizations} (model 1, 2 and 3) adopted in this study to place cosmology-independent constraints on $r_{\rm d}$ and $H_{0}$. The fourth case is a $\Lambda$CDM cosmological model.}
}
\label{table:models}
\end{table}
}

\subsection{Inference}\label{sect:inference}

{
We fit four models listed in Table~\ref{table:models} to observational data of type Ia supernovae, BAO and lensed quasars. Constrained model parameters include $r_{\rm d}$, $H_0$ and all remaining free polynomial coefficients (or density parameters in the case of $\Lambda$CDM model). The posterior distributions of the parameters are obtained using Affine-Invariant Monte Carlo Markov Chains (MCMC)  \citep{gw2010}, and in particular the python module \texttt{emcee} \citep{fm13}. For the sake of completeness, 
we also derive constraints on the deceleration parameter $q_0$ using the 
MCMC samples. Appendix \ref{Appendix:parametrizations} outlines the relations between polynomial coefficients, which are primary parameters in our fits, and $q_0$.
}

The likelihoods {of the distances measured from} lensed quasars are either given as a skewed log-normal distribution\footnote{{Full names and coordinates of each lens are given in the H0LiCOW~\textsc{XIII} paper \citep{Wong2019}}} (for B1608) or as {samples} of points {from the H0LiCOW model posteriors} (for RXJ1131, HE0435, PG1115, J1206 and WFI2033). The probability density is obtained by constructing a Gaussian kernel density estimator (KDE). {For the l}ens systems HE0435 and WFI2033 only a robust measurement of their time-delay distance\footnote{$D_{\Delta t}=(1+z_{\rm l})D_{\rm A,l}D_{\rm A,s}/D_{\rm A,ls}$ {, where $D_{\rm A,l}$, $D_{\rm A,s}$ and $D_{\rm A,ls}$} are the angular distance to the lens galaxy, lensed quasar and between the lens and the quasar; $z_{\rm l}$ is redshift of the lens galaxy} is provided, {which is the only robust distance currently derived from time-delay lensing in the presence of significant perturbers at lower redshift}. For the remaining four lenses {(B1608, RXJ1131, PG1115, J1206)}, information on both their time-delay distances and their angular diameter distances is available. {For the remaining observables (BAO, SNe, quasars, and SH0ES or CCHP), t}he general form of the likelihood for each data set is given by
\begin{eqnarray}
\nonumber \mathcal{L}\ &=&\ p(\mathrm{data|model})\ \propto e^{- \chi^2 / 2}\ ,\\
\chi^2 &=& \textbf{r}^{\dagger} \textbf{C}^{-1} \textbf{r},
\end{eqnarray}
where \textbf{C} is the covariance matrix of the data and \textbf{r} corresponds to the difference between the predicted and the observed values. The final likelihood is a product of the separate likelihoods corresponding to each data set. 

A uniform prior is used for the parameters, {for ease of comparison with previous work}. In particular, the value of $r_{\rm d}$ is kept between 0 and 200~Mpc and, if applicable, $\Omega_k$ between -1 and 1 and $\Omega_m$ between 0.05 and 0.5, to ensure consistency with the priors on $\Omega_m$ by H0LiCOW. {These priors
\adriano{do not skew the inference}, at least with the current uncertainties. The upper and lower boundaries of $r_{\rm d}$ do not influence any of the results.} \rev{For the coefficients of the expansion ($b_i$, $c_i$ and $d_i$ in Table \ref{table:models}), we used a uniform prior without limits.} {In all cases, best fit values are given by the posterior mean and errors provide 68.3 percent confidence intervals.} {The code to generate the results in this paper is publicly available on Github\footnote{\href{https://github.com/Nikki1510/cosmic_dissonance}{https://github.com/Nikki1510/cosmic$\_$dissonance}}.}

\nikki{\section{Comparison with the CMB: data and models} \label{sect:CMBcomp}}

\noindent {
The sound horizon and the Hubble constant are independently measured from the CMB. For the standard flat $\Lambda$CDM cosmological model, the Planck observations yield $r_{\rm d} = 147.2 \pm 0.3$ Mpc and $H_0 = 67.4 \pm 0.5$ km~s$^{-1}$~Mpc$^{-1}$ \citep{Planck2018}. As we shall demonstrate, both parameters are strongly discrepant with their counterparts determined from late-time observations. In the following subsections, we describe how we quantify this tension and outline a few popular extensions of the standard cosmological model devised to reduce the discrepancy.} \\

\subsection{Quantifying the tension}\label{sect:tension}
In order to check whether or not our results for $r_{\rm d}$ and $H_0$ are in agreement with those obtained by \textit{Planck}, the Gaussian odds indicator $\tau$ is used \citep{Verde2013,Ber2016}:
\begin{equation}
\tau = \frac{\int \Bar{P}_A \Bar{P}_B \; dx }{\int P_A P_B \; dx}.
\label{odds}
\end{equation}
Here, $P_{A}$ and $P_B$ denote the posterior distributions of experiments $A$ and $B$, while $\Bar{P}_{A}$ and $\Bar{P}_{B}$ correspond to the same distributions after a shift has been performed, such that the maxima of $P_A$ and $P_B$ coincide. A high value for $\tau$ means that it is unlikely that both experiments measure the same quantity. In {an} idealized situation, when experiment $A$ yields a measurement with infinite precision ($P_{A}$ is given a $\delta$ function), the odds indicator equals the ratio of probability $P_B$ evaluated at best fit values returned by both experiments. Eq.~\ref{odds} generalizes this interpretation to cases where both measurements have non-zero uncertainties.

A more intuitive scale representing the discrepancy between two measurements is a number-of-sigma tension, {and it} can be directly derived from the odds ratios {\citep[see e.g.][]{Ber2016}}. First, the odds indicator is used to calculate the probability enclosed by a contour $r$ such that $1 / \tau = e^{- \frac{1}{2} r^2} $. The probability is then converted to a number of sigma tension, using a one-dimensional cumulant (the error function).

\begin{figure}
	\centering
	\includegraphics[width=\linewidth]{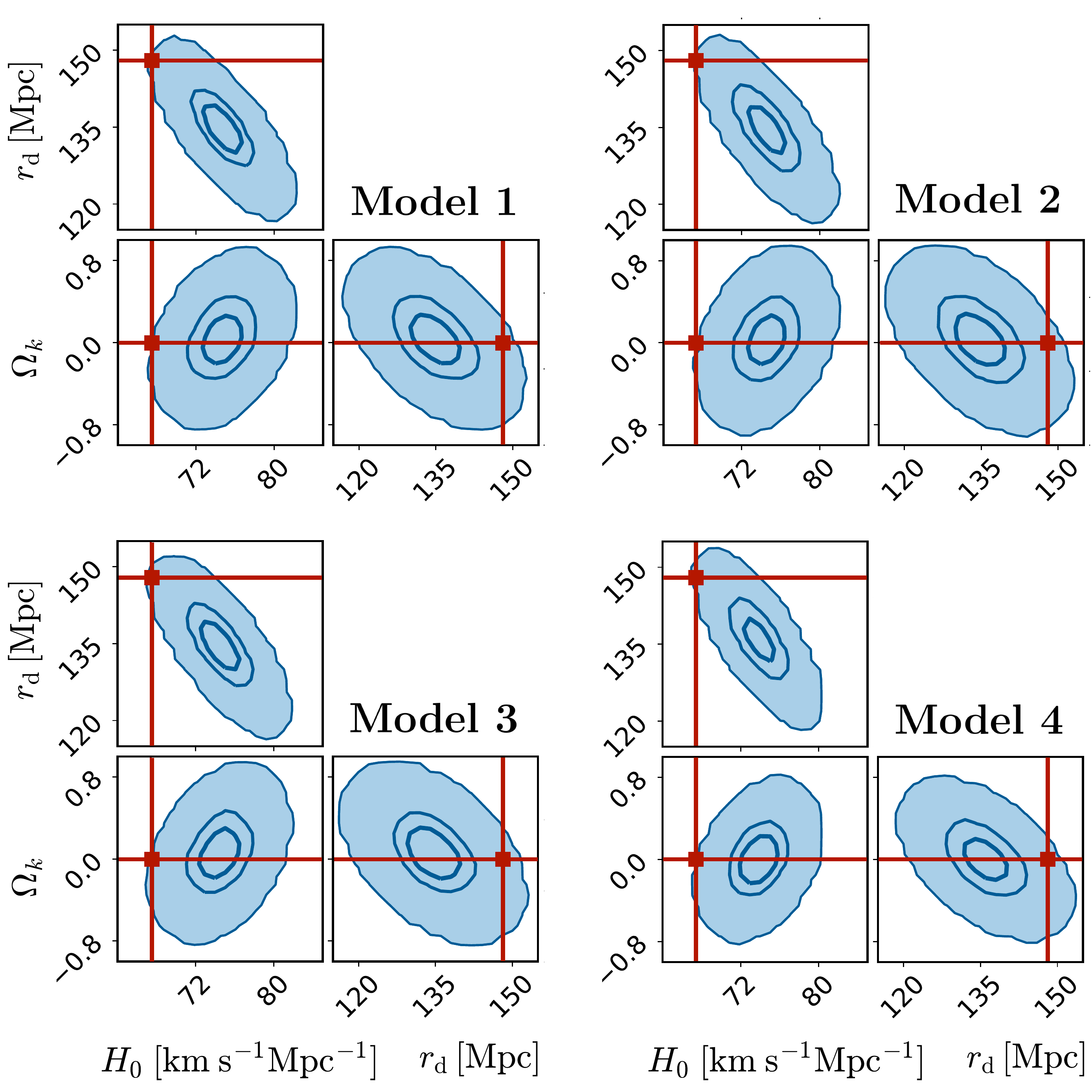}
	\caption{{
	Constraints on the sound horizon $r_{\rm d}$, the Hubble constant $H_0$ and $\Omega_k$ from late-time observations including BAO (BOSS), type Ia supernovae (Pantheon), gravitational lensing (H0LiCOW) and the cosmic distance ladder calibrated with Cepheids (SH0ES). The panels show results for three cosmology-independent models listed in Table~\ref{table:models} and a $\Lambda$CDM cosmological model. The red lines indicate the best fit values obtained from Planck for a flat $\Lambda$CDM cosmological model. The contours indicate 1-, 2- and 5$\sigma$ confidence regions of the posterior probability (the latter obtained by Gaussian extrapolation). All panels demonstrate a 5$\sigma$ tension between $r_{\rm d}$ and $H_0$ measured from the CMB and the late-time observations.
	}}
	\label{fig:4models}
\end{figure}

\begin{figure}
	\centering
	\includegraphics[width=\linewidth]{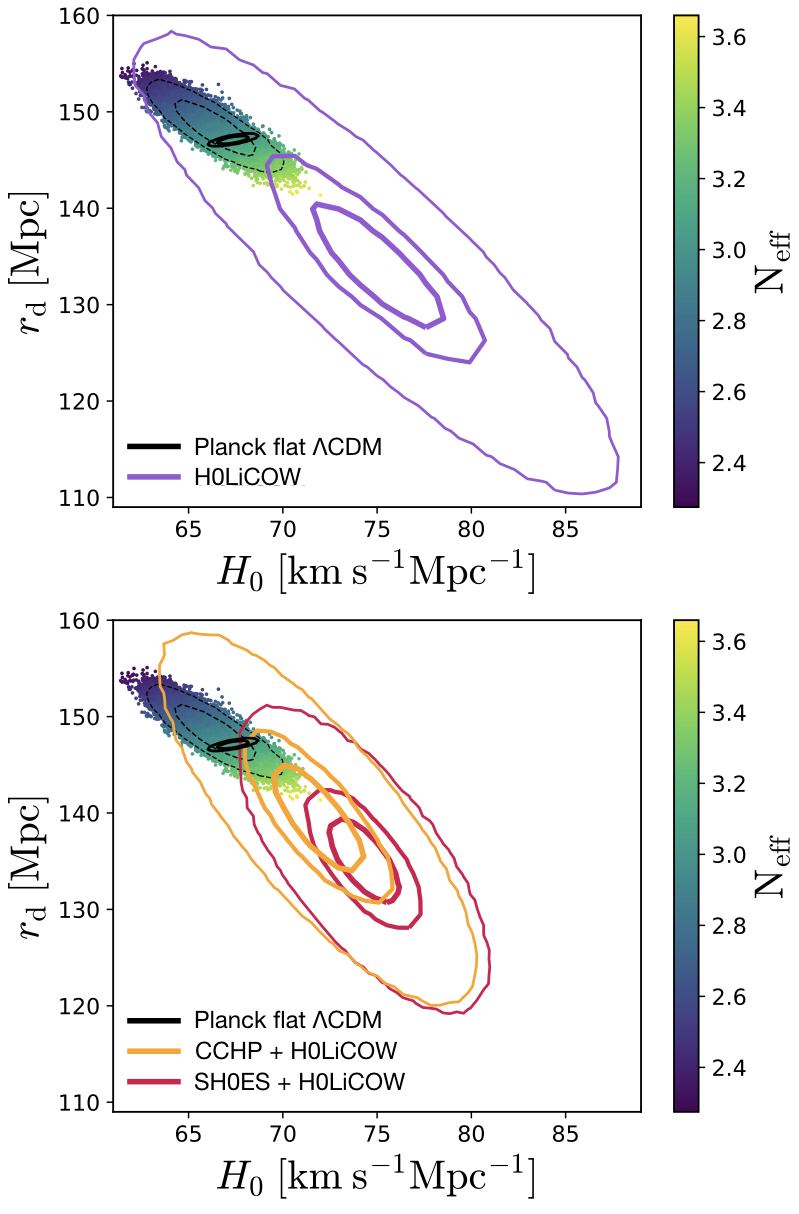}
	\caption{{
	Comparison between the sound horizon $r_{\rm d}$ and the Hubble constant $H_0$ measured from \textit{Planck} observations of the CMB (assuming a flat $\Lambda$CDM) and late-time observations (using flat model 3) \nikki{obtained by calibrating SN and BAO measurements with} three different absolute distance calibrations from: gravitational lensing (H0LiCOW), the cosmic distance ladder with Cepheids (SH0ES) or the TRGB (CCHP). For the late-time data, the contours show 1-, 2- and 5$\sigma$ confidence regions of the posterior probability (the latter obtained by Gaussian extrapolation). The Planck constraints (1- and 2$\sigma$ confidence regions) are obtained for the standard effective number of neutrinos (black solid line) and a model with a free effective number of neutrinos (black dashed lines, color points).
	}}
	\label{fig:Neff}
\end{figure}

\subsection{{Extensions of the $\Lambda$CDM model}}
\label{sect:extensions}
Any tension between late-time measurements and CMB-based model-dependent inference may be caused by unknown systematics, or it can mean that our knowledge of the physics underlying the expansion history is incomplete. The standard flat $\Lambda$CDM model can be extended by either changing physics in the early Universe \nikki{(pre-recombination; this will be referred to as \textit{early-time} modifications)} or at later epochs \nikki{(post-recombination; this will be referred to as \textit{late-time} modifications)}. In the first case, one can decrease the sound horizon inferred from the CMB observations by 
adding \rev{an energy-momentum tensor beyond the standard model}, which effectively increases $H(z)$ in the early Universe. In order to keep the observed angular scales imprinted in the CMB unchanged, this alteration automatically implies an increase in {the value of} $H_{0}$. Therefore, the overall effect of early-time modifications is a shift of both $r_{\rm d}$ and $H_{0}$ towards the measurements from late-time observations. \nikki{In the second approach, one may obtain higher values of $H_0$ by decreasing the expansion rate at intermediate redshifts. This can be done by modifying the dark energy density such that it increases over time. } Although many late-time extensions of the standard model can quite easily increase $H_{0}$ inferred from the CMB, $r_{\rm d}$ cannot be modified as appreciably as $H_{0}$ -- as it is primarily driven by physics in the early Universe.

In order to explore different resolutions of the tension in $H_0$ and $r_{\rm d}$ on the grounds of new physics, we consider several extensions of the standard $\Lambda$CDM model. \adriano{Although t}he selected models do not exhaust all possible proposals from the literature, they are sufficiently representative in terms of covering most possible model-dependent alterations of $H_{0}$ and $r_{d}$ inferred from the CMB. In what follows, the inference for early dark energy and PEDE (described below) have been obtained using a \textit{Planck} compressed likelihood, as detailed in Appendix \ref{appendix}. {For the remaining models, we use publicly available MCMC chains (based on Planck's temperature and polarization data) from the Planck Legacy Archive\footnote{\href{https://pla.esac.esa.int}{https://pla.esac.esa.int}} \citep{Planck2018}.}

\subsubsection{{\nikki{Early-time (pre-recombination) extensions}}}

$\bullet \,$ \textit{Effective Number of Relativistic Species ($\textrm{N}_{\textrm{eff}}$)}. \\
\noindent \nikki{In this extension of $\Lambda$CDM, there are additional relativistic particles} that contribute to the radiation density of the early Universe, resulting in  $\textrm{N}_{\textrm{eff}}> 3.$ An increased radiation density leads to a later matter-radiation equality and to an increased expansion rate in the early Universe, {leaving an observational imprint on the CMB \citep{Eis2004,Hann2003,Mor2018}}. 
This in turn reduces the value of the sound horizon $r_{\rm d}$ at recombination \rev{and increases $H_{0}$ derived from the CMB}, thereby relieving some of the tension between late-time and CMB measurements \citep{Carneiro2019,Gel19}. \\

\noindent $\bullet \,$ \textit{Early Dark Energy}. \\
\noindent The expansion rate in the early Universe could also be increased by the presence of a more general form of dark energy. This additional dark energy should have a noticeable contribution to the energy budget at high redshifts, but should dilute away faster than radiation to leave the evolution of the Universe {after recombination} unchanged \citep{Dor2007,Lin2008}. As a promising example of this class of models, we consider early dark energy which behaves nominally as a scalar field $\phi$ with a potential $V(\phi)\propto[1-\cos(\phi/f)]^{3}$ \citep{Poulin2019}. In the effective fluid description, the energy density ${\rho_{EDE}}$ evolves as 
\begin{equation}
    \rho_{\rm EDE}(a)=\frac{2\rho_{EDE}(a_{\rm c})}{1+(a/a_{\rm c})^{9/2}}
\end{equation} 
with the scale factor $a$ \citep{Poulin2018}. {The early dark energy equation of state approaches asymptotically $-1$ for $a\ll a_{c}$ and $1/2$ for $a\gg a_{c}$.} When fitting the model to the CMB data, we adopt the following flat priors in $\log_{10}(a_{\rm c})$ and $f_{\rm EDE}=\Omega_{\phi}(a_{\rm c})/\Omega_{\rm tot}(a_{\rm c})$: $-4.0<\log_{10}(a_{\rm c})<-3.2$ and $0.1>f_{\rm EDE}>0$.

\subsubsection{{\nikki{Late-time (post-recombination) extensions}}}

$\bullet \,$ \textit{Time-dependent dark energy (wCDM)}. \\
\noindent The $w$CDM cosmology introduces the equation of state parameter $w$ as a free parameter (as opposed to the fixed $\Lambda$CDM value of $w = -1$), so that the dark energy density ${\rho_{DE}}$ can change as a function of redshift as
\begin{equation}
{\rho_{DE} (z) = \rho_{DE,0} (1+z)^{3(1+w)}.}
\end{equation}

\noindent $\bullet \,$ \textit{Phenomenologically Emergent Dark Energy (PEDE)}. \\
\noindent In the PEDE model, dark energy has no effective role in the early Universe but emerges at later times \citep{Li2019b}. The redshift evolution of the dark energy density is described by
\begin{equation}
{\rho_{DE} (z)} = {\rho_{DE,0}} \times {[1 - \tanh(\log_{10}(1 + z))]},
\label{PEDE}
\end{equation}
giving it the same number of degrees of freedom as $\Lambda$CDM. We emphasize that this parameterization is mostly \textit{ad hoc}.

\section{Results and Discussion}
\label{sect:results}

The values of the sound horizon and other parameters inferred from the six lenses, Pantheon SN sample and BAO measurements (BOSS) using three models that employ polynomial parametrization or a $\Lambda$CDM model are listed in  Table~\ref{table1}. The tension with \textit{Planck} flat $\Lambda$CDM and late-time extension models is displayed in the last rows and ranges from 2$\sigma$ to 3$\sigma$. When combining the distance calibration from the lensed quasars with that from SH0ES (the distance ladder with Cepheids), the constraints on $r_{\rm d}$ are tighter and the tension with \textit{Planck} increases to 5$\sigma$, as can be seen in Table~\ref{table2}. \nikki{The corresponding Bayesian Information Criterion (BIC) values are the lowest for model 4 ($\Lambda$CDM). {However, the differences in BIC do not exceed \rev{6} (substantial level on the Jeffreys scale), with a minimum of 1 for model 1 (barely worth mentioning level on the Jeffreys scale).}} Figure~\ref{fig:4models} compares constraints on $H_0$, $r_{\rm d}$ and $\Omega_{k}$ from late-time observations including the prior from SH0ES to the best fit parameters derived from Planck assuming a flat $\Lambda$CDM model. For all models, the \textit{Planck} parameters lie on the 5$\sigma$ contour in the $H_0 - r_{\rm d}$ plane, demonstrating that the tension is independent of the chosen expansion family.

\nikki{In Table \ref{table3}, some other combinations of data sets have been explored. This includes a calibration of lenses + CCHP instead of SH0ES, inclusion of several volume-averaged and Ly-$\alpha$ BAO and the addition of high redshift quasars as secondary standard candles.} {Considering all results based on the main  data sets (H0LiCOW, SN, BAO/BOSS) with the cosmic distance ladder (SH0ES or CCHP), we find $r_{\rm d}= (137 \pm 3^{stat.} \pm 2^{syst.})$ Mpc, where the systematic error accounts for differences between SH0ES and CCHP distance calibration. In addition, we run an inference free of any SN data, thus only using lensed quasars and BAO measurements from BOSS, $D_V$ and Ly-$\alpha$ with a flat $\Lambda$CDM model.\footnote{{For the flat $\Lambda$CDM model, we adopted a prior of $\Omega_M = \mathcal{U}[0.05, 0.5].$}} This results in the following values for the cosmological parameters:
$r_{\rm d} = 138.6 \pm 3.8 \; \rm{Mpc}$, 
$H_0 r_{\rm d} = 10166 \pm 142 \; \rm{km}\, \rm{s}^{-1}$,
$\Omega_m = 0.29 \pm 0.02$. \\
}

\begin{table*}
\caption{{Posterior mean and standard deviation for the sound horizon $r_{\rm d}$, $H_{0}r_{\rm d}$ and $q_{0}$ inferred from late-time observations including H0LiCOW lensing observations, Pantheon SN sample and BAO measurements (BOSS). The fit quality is summarized in terms of log-likelihood at the maximum posterior probability, $\ln\mathcal{L}_{m.a.p.}$, and the Bayesian Information Criterion $\mathrm{BIC}\ =\ \ln(N)k-2\ln(\mathcal{L}_{m.a.p.})$, where $N$ is the number of data points and $k$ is the number of free parameters. The odds indicator $\tau$ quantifies the tension between $r_{\rm d}$ and $H_0$ measured from late-time observations and the \textit{Planck} data (for the standard flat $\Lambda$CDM model and its two extensions with a free effective number of neutrinos or early dark energy).
}}
\centering          
\begin{tabular}{l l l l l}
\hline 

 & \multicolumn{4}{c}{flat ($\Omega_{k}=0$)} \\
parameter &  model 1 & \nikki{model 2} & \nikki{model 3} & model 4 (f$\Lambda$CDM)\\ 
\hline
\\
$r_{\rm d}$ (Mpc)  & 132.7 $\pm$ 4.2 & 132.9 $\pm$ 4.4 & 134.2 $\pm$ 4.4 &  136.9 $\pm$ 3.7 \\
$H_{0}r_{\rm d}$ (km~s$^{-1}$)  & 10107 $\pm$ 147 & 10065 $\pm$  150  & 10052 $\pm$ 152 & 10038 $\pm$ 136 \\ 
$q_{0}$  & -0.7 $\pm$ 0.07 & -0.5 $\pm$ 0.2 & -0.4 $\pm$ 0.3  &  \rev{-0.55 $\pm$ 0.03}\\
$\ln\mathcal{L}_{m.a.p.}$  & -86.3 & -86.1 & -86.7 &  -87.7 \\
BIC score  & 193 & 196 & 198 & 192 \\
$\ln\tau$ (Planck $\Lambda$CDM) & 6.6 (3.2$\sigma$) & 5.7 (2.9$\sigma$) & 5.0 (2.7$\sigma$) & 5.7 (2.9$\sigma$) \\ 
$\ln\tau$ (Planck $\Lambda$CDM+N$_{\rm eff}$)  & 6.3 (3.1$\sigma$) & 5.6 (2.9$\sigma$) & 4.9 (2.7$\sigma$)  & 5.0 (2.7$\sigma$) \\
$\ln\tau$ (Planck early DE)  & 5.1 (2.8$\sigma$) & 4.4 (2.5$\sigma$) & 3.7 (2.3$\sigma$) & 3.7 (2.2$\sigma$) \\
\hline\hline
 & \multicolumn{4}{c}{free $\Omega_{k}$} \\
parameter &  model 1 & \nikki{model 2} & \nikki{model 3} & model 4 ($\Lambda$CDM)\\ 
\hline
\\
$r_{\rm d}$ (Mpc) & 129.2 $\pm$ 5.7 & 130.6 $\pm$ 5.9 & 131.2 $\pm$ 6.1 & 137.2 $\pm$ 4.8 \\
$H_{0}r_{\rm d}$ (km~s$^{-1}$)  & 10045 $\pm$ 155 & 10033 $\pm$ 157  & 10017 $\pm$ 160 & 10041 $\pm$ 156 \\
$\Omega_{k}$  & 0.18 $\pm$ 0.2 & 0.13 $\pm$ 0.2 & 0.15 $\pm$ 0.2 & -0.01 $\pm$ 0.2  \\
$q_{0}$  & -0.6 $\pm$ 0.1 & -0.4 $\pm$ 0.2 & -0.4 $\pm$ 0.3 &  \rev{-0.56 $\pm$ 0.07}\\
$\ln\mathcal{L}_{m.a.p.}$  & -86.1 & -85.9 & -86.4 & -87.7 \\
BIC score  & 196 & 200 & 201 & 196 \\
$\ln\tau$ (Planck $\Lambda$CDM)  & 5.6 (2.9$\sigma$) & 4.6 (2.6$\sigma$)  & 4.0 (2.4$\sigma$) & 4.2 (2.4$\sigma$)\\
$\ln\tau$ (Planck $\Lambda$CDM+N$_{\rm eff}$)  & 5.7 (2.9$\sigma$) & 4.7 (2.6$\sigma$) & 4.2 (2.4$\sigma$) & 3.9 (2.3$\sigma$) \\
$\ln\tau$ (Planck early DE)  & 4.5 (2.6$\sigma$) & 3.6 (2.2$\sigma$) & 3.1 (2.0$\sigma$) & 2.6 (1.8$\sigma$) \\
\hline
\end{tabular}
\label{table1}
\end{table*}

\begin{table*}
\caption{
{The same as Table~\ref{table1}, but for fits based on the H0LiCOW lensing, Pantheon SN sample, BAO measurements (BOSS) and $H_0$ from SH0ES.}
}
\centering    
\begin{tabular}{l l l l l}
\hline 

 & \multicolumn{4}{c}{flat ($\Omega_{k}=0$)} \\
parameter &  model 1 & \nikki{model 2} & \nikki{model 3} & model 4 (f$\Lambda$CDM)\\ 
\hline
\\
$r_{\rm d}$ (Mpc) & 135.1 $\pm$ 2.8 & 135.0 $\pm$ 2.9 & 135.1 $\pm$ 2.9 & 136.1 $\pm$ 2.7 \\
$H_{0}r_{\rm d}$ (km~s$^{-1}$)  & 10079 $\pm$ 143 & 10055 $\pm$ 148 & 10038 $\pm$ 153 & 10037 $\pm$ 136 \\ 
$q_{0}$  & -0.6 $\pm$ 0.07 & -0.4 $\pm$ 0.2 & -0.4 $\pm$ 0.3  &  \rev{-0.55 $\pm$ 0.03}\\
$\ln\mathcal{L}_{m.a.p.}$  & -86.6 & -86.4 & -86.8 &  -87.7 \\
BIC score  & 193 & 197 & 198 & 192 \\
$\ln\tau$ (Planck $\Lambda$CDM)  & 15.1 (5.1$\sigma$) & 15.0 (5.1$\sigma$) & 13.9 (4.9$\sigma$) & 15.1 (5.1$\sigma$) \\
$\ln\tau$ (Planck $\Lambda$CDM+N$_{\rm eff}$)  & 9.9 (4.1$\sigma$)  & 9.7 (4.0$\sigma$) & 9.2 (3.9$\sigma$) & 9.1 (3.9$\sigma$) \\
$\ln\tau$ (Planck early DE)  & 9.4 (3.9$\sigma$) & 9.2 (3.9$\sigma$) & 8.6 (3.7$\sigma$) & 8.7 (3.8$\sigma$) \\
\hline\hline
 & \multicolumn{4}{c}{free $\Omega_{k}$} \\
parameter &  model 1 & \nikki{model 2} & \nikki{model 3} & model 4 ($\Lambda$CDM)\\ 
\hline
\\
$r_{\rm d}$ (Mpc)  & 134.8 $\pm$ 3.2 & 134.7 $\pm$ 3.3 & 134.6 $\pm$ 3.3 &  136.1 $\pm$ 3.2 \\
$H_{0}r_{\rm d}$ (km~s$^{-1}$) & 10067 $\pm$ 156 & 10042 $\pm$ 161 & 10021 $\pm$ 161 & 10035 $\pm$ 152 \\
$\Omega_{k}$  & 0.04 $\pm$ 0.2 & 0.03 $\pm$ 0.2 & 0.06 $\pm$ 0.2 & 0.003 $\pm$ 0.2 \\
$q_{0}$  & -0.6 $\pm$ 0.09 & -0.4 $\pm$ 0.2 & -0.4 $\pm$ 0.3 &  \rev{-0.55 $\pm$ 0.07}\\
$\ln\mathcal{L}_{m.a.p.}$  & -86.7 & -86.5 & -86.8 & -87.7 \\
BIC score  & 198 & 201 & 202 & 196 \\
$\ln\tau$ (Planck $\Lambda$CDM)  & 13.3 (4.8$\sigma$) & 13.2 (4.8$\sigma$) & 12.7 (4.7$\sigma$) & 12.8 (4.7 $\sigma$) \\
$\ln\tau$ (Planck $\Lambda$CDM+N$_{\rm eff}$)  & 9.2 (3.9$\sigma$) & 9.0 (3.9$\sigma$) & 8.9 (3.8$\sigma$) & 8.2 (3.6$\sigma$) \\
$\ln\tau$ (Planck early DE)  & 8.3 (3.7$\sigma$) & 8.2 (3.7$\sigma$) & 8.0 (3.6$\sigma$) & 7.3 (3.4$\sigma$) \\
\hline
\end{tabular}

\label{table2}  
\end{table*}

\subsection{\nikki{Early-time} extensions}

A possible solution for the tension is an extension to the early Universe physics, such as an additional component of relativistic species. \textit{Planck} 2018 chains with free $\textrm{N}_{\textrm{eff}}$ {(based on full temperature and polarization data)} have been used to investigate this scenario. In Figure~\ref{fig:Neff}, \textit{Planck} + free $\textrm{N}_{\textrm{eff}}$ is compared to results from model 3 using SN + BAO with only the H0LiCOW lenses as calibrator (upper panel) and using a combination of H0LiCOW lenses and either SH0ES or CCHP as calibrators (lower panel). A higher value of $\textrm{N}_{\textrm{eff}}$ is shown to move the \textit{Planck} value to a lower $r_{\rm d}$ and a higher $H_0$, therefore alleviating the tension to some extent. {In this case, the combined analysis of \textit{Planck} and low-redshift data yields $\rm{N}_{\rm eff}=3.24\pm0.16$.} This effect is only convincing when the late-time measurements are calibrated with H0LiCOW and CCHP, since the alternative Cepheid calibration is still in tension with the \textit{Planck}$+\textrm{N}_{\textrm{eff}}$ extension (see Table~\ref{table2}).

\begin{figure}
	\centering
	\includegraphics[width=\linewidth]{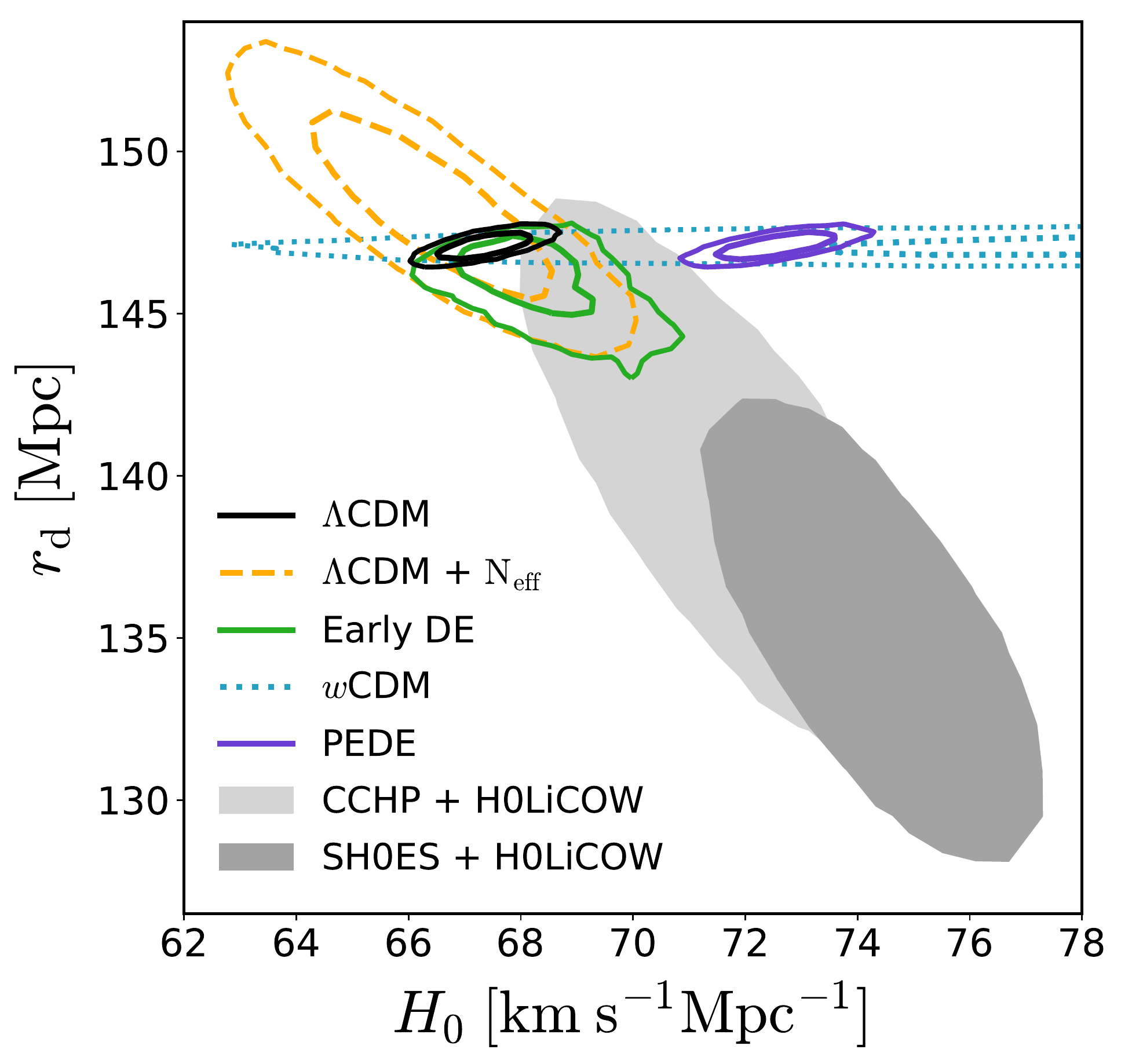}
	\caption{\nikki{The effect of four different extensions of the flat $\Lambda$CDM model on the sound horizon and the Hubble constant measured from the \textit{Planck} data. The models considered here are $\Lambda$CDM + free $\textrm{N}_{\textrm{eff}}$, early dark energy, wCDM and PEDE. The CMB-based constraints are compared to the measurements from late-time observations (SN + BAO + H0LiCOW + SH0ES/CCHP) shown with the gray shaded contours. The late-time measurements are obtained with model 3 (see Table \ref{table:models}) and show the $2\sigma$ credibility regions. }}
	\label{fig:extensions}
\end{figure}

\subsection{Tension between the CMB and late-time observations}
{Figure~\ref{fig:extensions} demonstrates the potential of the selected extensions of the standard $\Lambda$CDM model outlined in Section~\ref{sect:extensions} to resolve the tension between $r_{\rm d}$ and $H_0$ measured from the CMB and late-time observations.} 
 \nikki{The shaded gray contours show constraints from late-time observations using model 3 with $\Omega_k = 0$. Thanks to using a polynomial parametrization, these measurements are marginalized over a wide class of the expansion history and in this sense they are independent of cosmological model. We show results for distance calibrations based on the H0LiCOW lenses combined with SH0ES or {CCHP}. The contours in color show constraints from Planck for the flat $\Lambda$CDM model (black contours) and its four extension.}

As clearly seen from Figure~\ref{fig:extensions}, none of the $\Lambda$CDM extensions manage to convincingly unify the \textit{Planck} measurements with the late-time ones if the SH0ES calibration is used to anchor the distance ladder. In particular, late-time extensions involving different generalizations of the cosmological constant can increase the $H_0$ value inferred from the CMB, but they leave $r_{\rm d}$ unchanged. Although early-time extensions can potentially match both $H_{0}$ and $r_{\rm d}$ from low-redshift probes and the CMB, 
that this may happen by expanding the posterior probability contours rather than shifting the best fit values \citep[see also][]{Ber2016,Kar2016}, as demonstrated in Fig.~\ref{fig:extensions}. In this respect, both early dark energy models and extensions with extra relativistic species are quite similar.
 The apparent difference between their probability contours reflect differences in the priors. While a free effective number of 
relativistic species can either decrease or increase the 
sound horizon, early dark energy (with positive energy density) can only increase the energy budget, and thus decrease the sound horizon.

\begin{figure}
	\centering
	\includegraphics[width=\linewidth]{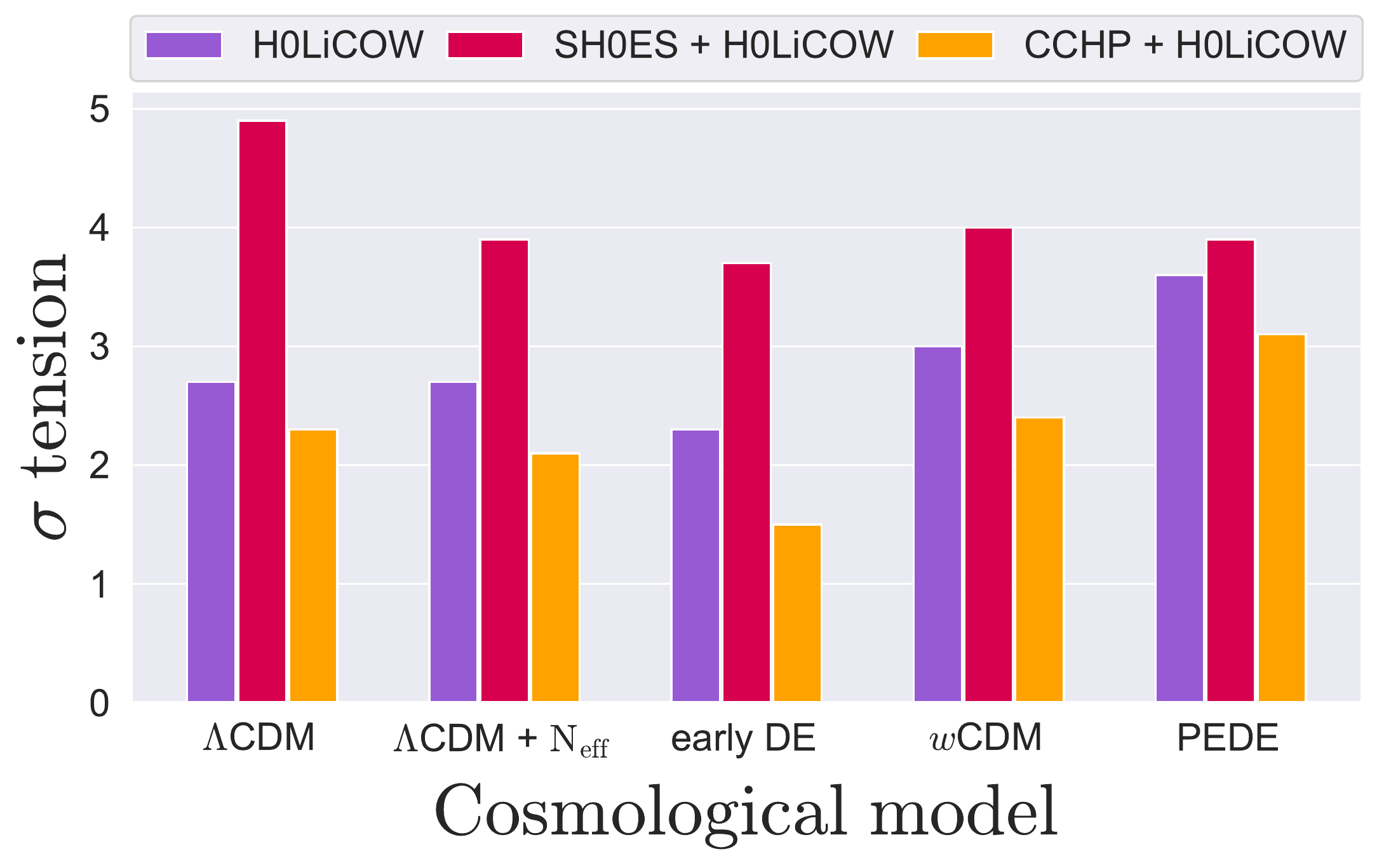}
	\caption{{
	Tension between the sound horizon and the Hubble constant measured from late-time observations and the CMB for the following cosmological models: $\Lambda$CDM, $\Lambda$CDM + $\textrm{N}_{\textrm{eff}}$, early DE, wCDM, PEDE-CDM (flatness assumed in all cases). Late-time observations include BAO, type Ia supernovae and three different absolute distance calibrations from gravitational lensing (H0LiCOW), the cosmic distance ladder with Cepheids (SH0ES) or the TRGB (CCHP).}
	}
	\label{fig:tension}
\end{figure}

Figure~\ref{fig:tension} summarizes the tension in the $H_0 - r_{\rm d}$ plane between late-time measurements and \textit{Planck} with different extensions of $\Lambda$CDM. 
To ensure a fair comparison, the same $\Lambda$CDM extensions are used in the late-time and CMB-based inference. Therefore, the \textit{Planck} PEDE-CDM results have been compared to late-time results obtained with PEDE-CDM, and the \textit{Planck} wCDM results to late-time results using \textit{w}CDM. For the early-time extensions this is not of great importance, since their effects do not influence the low-redshift measurements.

\nikki{By adopting different models of polynomial parametrizations (models 1, 2 and 3), we minimize the dependence on a cosmological model. Although our inference with these models does not depend on $\Lambda$CDM, it does have a weak dependency on General Relativity (GR).} The lensed quasars that are used to calibrate the distance ladder need GR in order to calculate the angular diameter distance, through the \textit{Ansatz} that the lensing potential (used in the time-delay inference) is exactly twice the gravitational potential (used to obtain $D_{A}\propto \mathrm{c}^{3}\Delta t/\sigma^{2}$ from stellar kinematics). {However, the role of this GR dependence is subdominant with current $D_{A}$ uncertainties ($10\%-20\%$).} {On the other hand, GR also} enters the early-Universe expansion through the `abundances' of different components ($\Omega_{m},$ $\Omega_{\rm{de}},$ $\rm{N}_{\rm{eff}}$).

\subsection{One lens at a time}
Since $H_0$ and $r_{\rm d}$ are constants, they must be independent of the chosen indicators. If they are inferred from each indicator separately, any trend will signal residual systematics, either in the indicators themselves or in the parameterization that is chosen to extrapolate $H(z)$ down to $H_0.$

The H0LiCOW collaboration have shown that, if $H_0$ is obtained from lenses in a flat-$\Lambda$CDM model, there is a weak trend in {its} inferred value versus lens redshift, with lower-redshift (resp. higher-redshift) lenses being more (resp. less) discrepant with the \textit{Planck} measurements \citep{Wong2019}. Even though this trend is currently not significant (given current uncertainties), it may be indicative of intrinsic systematics in the lensing inference, or in the way that time-delay distances are converted into $H_0$ values through a flat-$\Lambda$CDM parameterization. 

\begin{figure}
	\centering
	\includegraphics[width=\linewidth]{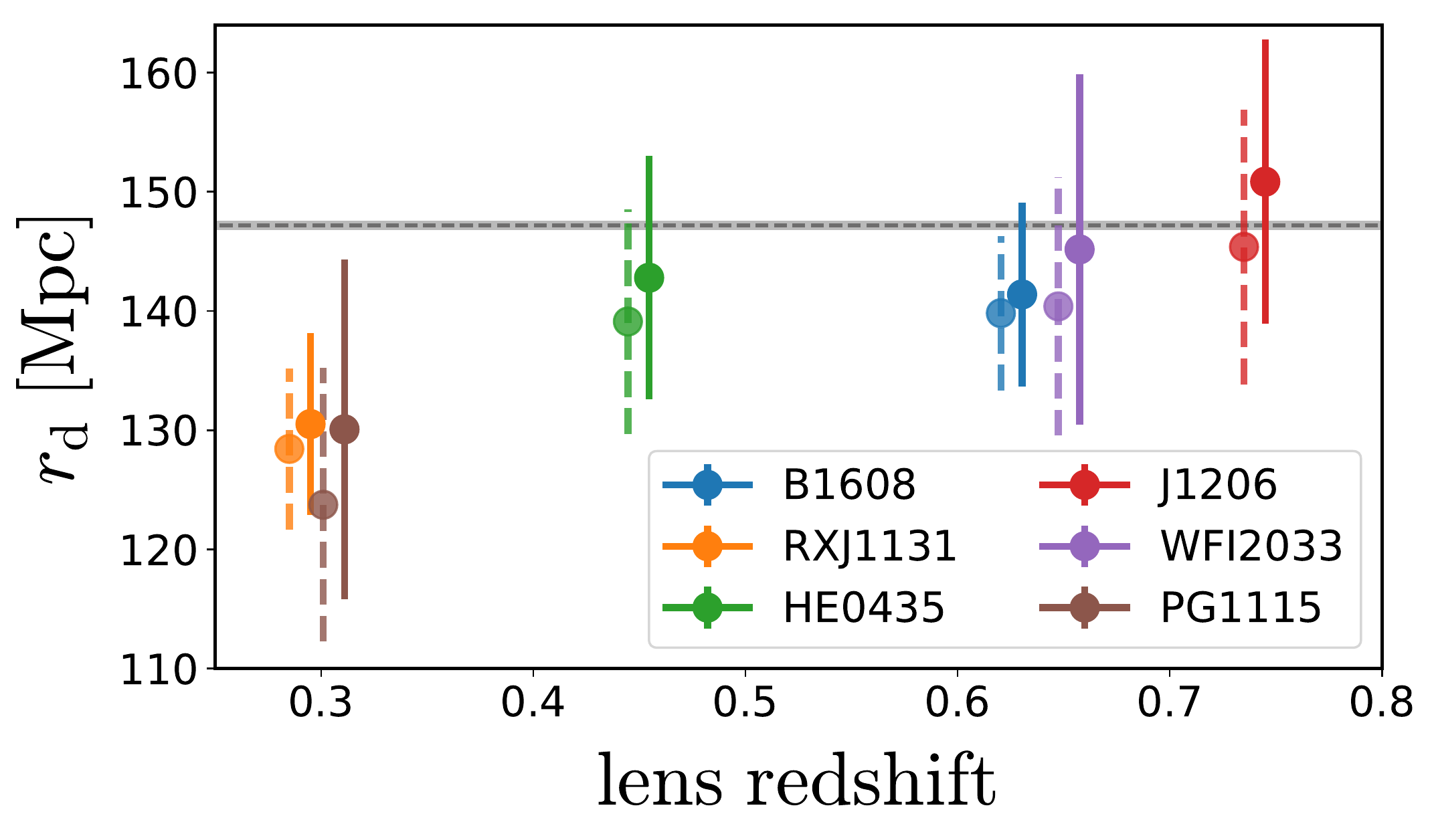}
\caption{{The sound horizon $r_{\rm d}$ measured from combining BAO and SNe data with H0LiCOW lensing observations of each lens separately. Here the distance calibration is set solely by the lensing observations of each individual lens. The measured sound horizon is shown as a function of lens redshift for fits with a flat model 3 (solid error bars) and a flat PEDE-CDM model (dashed error bars). For both models, the measurements show a slight trend of $r_{\rm d}$ increasing with lens redshift. The inference from models 1 and 2 is fully consistent with the model 3 results. The gray dashed line with shaded region shows \textit{Planck}'s value of $r_{\rm d}$ and its {(sub-percent)} uncertainty obtained for the standard flat $\Lambda$CDM model.}}
	\label{fig:separatelenses}
\end{figure}

{Here we repeat this test using more general models of the expansion history, specifically flat model 3 and flat PEDE-CDM model. Figure~\ref{fig:separatelenses} shows the sound horizon $r_{\rm d}$ measured from combining BAO and SNe data with lensing constraints from each lens separately. The results demonstrate that the distance calibration from H0LiCOW lenses shows a similar trend with lens redshift as the one shown by \cite{Wong2019} for a flat $\Lambda$CDM cosmology}. \adriano{Based on the sample-wide analysis by \citet{Millon20}, this weak trend cannot be explained simply on the basis of known systematics in the lens models or kinematics of each lens. We should emphasize, however, that this trend is \textit{not} statistically significant ($1.6\sigma$) yet.}

{Although the current weak trend of $r_{\rm d}$ with redshift of gravitatonal lens is consistent with being a statistical fluke, it is instructive to investigate if} there any expansion models that can re-absorb this (weak) trend. For example, a recent ($z\approx0.4$) change in dark energy may produce this behaviour, if the data are interpreted with expansion histories that are `too' smooth. For this reason, we examine the same lens-by-lens determination within the PEDE model family. The results are shown as dotted error-bars in Figure~\ref{fig:separatelenses}. Even the PEDE model with accelerated late-time expansion cannot eliminate the (weak) trend in $r_{\rm d}.$ The constraints set by the relative distance moduli of SN enforce PEDE to closely resemble the $\Lambda$CDM case, but with a higher matter content ($\Omega_{m}\approx0.345$) and smaller sound horizon ($r_{d}\approx138$~Mpc). Therefore, PEDE does not resolve the current tension.

\begin{table*}
\caption{{The same as Table~\ref{table1}, but for various combinations of late-time observations including two local determinations of $H_0$ (SH0ES or CCHP), measurements of isotropic BAO ($D_{\rm V}$) and anisotropic BAO from the Lyman-$\alpha$ forest of quasars (Ly-$\alpha$), and estimates of distance moduli from high-redshift quasars. The parameters are determined using model 3 with $\Omega_k=0$.}}
\centering          
\begin{tabularx}{\textwidth}{l C C C C}
\hline 
 & \multicolumn{4}{c}{flat ($\Omega_{k}=0$)} \\
parameter \hspace{1.5cm} &  \nikki{CCHP + H0LiCOW + SN + BAO (BOSS)} & \nikki{SH0ES + H0LiCOW + SN + BAO (BOSS + $D_{\rm V}$ + Ly-$\alpha$)}  & \nikki{H0LiCOW + SN + BAO (BOSS + $D_{\rm V}$ + Ly-$\alpha$)} & \nikki{SH0ES + H0LiCOW + SN + BAO (BOSS) + high-\textit{z} quasars} \\ 
\hline
\end{tabularx}
\begin{tabularx}{\textwidth}{l X X X X}
$r_{\rm d}$ (Mpc) & 139.5 $\pm$ 3.6 & 138.1 $\pm$ 2.7 & 138.6 $\pm$ 3.8 & 134.0 $\pm$ 2.8 \\
$H_{0}r_{\rm d}$ (km~s$^{-1}$)  & 10019 $\pm$ 152 & 10197.1 $\pm$ 135  & 10191 $\pm$ 138 & 10011 $\pm$ 149 \\ 
$q_{0}$  & -0.4 $\pm$ 0.4 & -0.9 $\pm$ 0.3 & -0.8 $\pm$ 0.3 & -0.2 $\pm$ 0.3 \\
$\ln\tau$ (Planck $\Lambda$CDM)  & 3.8 (2.3$\sigma$) & 12.8 (4.7$\sigma$) & 4.4 (2.5$\sigma$) & 17.1 (5.5$\sigma$) \\ 
$\ln\tau$ (Planck $\Lambda$CDM+N$_{\rm eff}$)  & 3.4 (2.1$\sigma$) & 8.0 (3.6$\sigma$) & 4.2 (2.4$\sigma$) & 11.0 (4.3$\sigma$) \\
$\ln\tau$ (Planck early DE)  & 2.1 (1.5$\sigma$) & 7.2 (3.4$\sigma$) & 2.8 (1.9$\sigma$) & 10.8 (4.3$\sigma$) \\
\hline
\end{tabularx}
\label{table3}  
\end{table*}

\section{Conclusions and Outlook}
\label{sect:conclusions}

We have combined the newest available low-redshift probes to obtain an estimate of the sound horizon \nikki{at the drag epoch, $r_{\rm d}$}. {In order to minimize the dependence on a cosmological model, we have used a set of polynomial parametrizations that are almost entirely independent of the underlying cosmology, as well as the standard $\Lambda$CDM model.} In the $H_0 - r_{\rm d}$ plane, we have found a tension of $5 \sigma$ between \textit{Planck} results using flat $\Lambda$CDM and late-time observations calibrated with H0LiCOW lenses and SH0ES. This tension reduces to {2.4$\sigma$} if CCHP results are used as a distance-ladder anchor instead of SH0ES. We have investigated whether early- or late-time extensions to the standard $\Lambda$CDM model can resolve the tension and examined models with free $\textrm{N}_{\textrm{eff}},$ early dark energy,  \textit{w}CDM and PEDE-CDM. None of these model extensions provide a satisfying solution to the Hubble tension problem {\citep[see also][]{Ayl2019,Kno2019}}, except for free $\textrm{N}_{\textrm{eff}}$ or early dark energy in combination with low redshift data calibrated by {CCHP} + H0LiCOW.

These findings can indicate that: (1) extensions of early-time physics are necessary; and/or (2) that systematics from different late-time probes are becoming comparable to the statistical uncertainties. Arguments based on local under-densities or peculiar velocities cannot resolve the tension: 
the $\approx3\sigma$ tension persists if the inverse-distance ladder is restricted to $z\geq0.03,$ where the role of peculiar velocities is $\lesssim0.1\%$ \citep[see also][]{Woj2019}. {Multiple secondary sources of errors in redshift measurements were studied by \citet{Davis2019}, but none of them seem to have any noticeable effect.} Another explanation may be that the standardization of SNe \textsc{I}a is not properly understood yet {\citep[as a caveat see e.g.][also Khetan et al. 2020]{Rig2015}}, or that there is some (hitherto undiscovered) source of systematics in \nikki{one of the other used data sets.} {If all astrophysical systematics are exhausted, one can also consider proposals involving non-standard physics in the local Universe such as screened fifth forces, which may bias $H_{0}$ measurements high via \rev{modulation of gravity-dependent pulsation periods of Cepheids} \citep[for more details see][]{Des2019}.
} For these reasons, we also provide a measurement that relies only on lenses and BAO, without any additional constraint from SNe, \nikki{in section} \ref{sect:results}.

The weak trend in Figure \ref{fig:separatelenses} may indicate residual systematics in the lens models, or the need for different low-\textit{z} expansion models, or it may vanish entirely with larger lens samples.
In order to check the robustness of the trend, {cosmography-grade models of more lenses are needed, over the whole $0.3\lesssim z\lesssim0.7$ current redshift interval and beyond.} Finally, the role of systematics in the lens mass models can be assessed once high-S/N spatially-resolved kinematics are available \citep{STA18, Yil19}, which would  {enable more flexible dynamical models} than the ones used so far on aperture-averaged velocity dispersions.

As a final remark, we emphasize that resolving the $H_0$ tension alone is not sufficient, since different models that can shift this value are still at tension with the inferred $r_{\rm d}$ from BAO and low-redshift indicators. {Also, a direct combination of the inference from late-time and CMB-based measurements that may be at $>3\sigma$ tension, hence hardly compatible with one another, should be justified.} Therefore, any new proposal to resolve the discrepancy between CMB-based and late-time measurements should consider both $H_0$ and $r_{\rm d},$ and examine the \textit{separate} inference upon \nikki{late-time and CMB-based} data.

\begin{acknowledgements}
We thank Chiara Spiniello for useful comments on an earlier version of this manuscript, and Inh Jee and Eiichiro Komatsu for discussions on the lensing distance likelihoods.
These time delay cosmography observations are associated with programs HST-GO-9375, HST-GO-9744, HST-GO-10158, HST-GO-12889, and HST-14254. Support for programs HST-GO-10158 HST-GO-12889 HST-14254 was provided to members of our team by NASA through grants from the Space Telescope Science Institute, which is operated by the Association of Universities for Research in Astronomy, Inc., under NASA contract NAS 5-26555. 
AA and RJW were supported by a grant from VILLUM FONDEN (project number 16599). This project is partially funded by the Danish council for independent research under the project ``Fundamentals of Dark Matter Structures'', DFF--6108-00470.
 This project has received funding from the European Research Council (ERC) under the EU’s Horizon 2020 research and innovation programme (COSMICLENS; grant agreement No. 787886) and from the Swiss National Science Foundation (SNSF).
 This work was supported by World Premier International Research Center Initiative (WPI Initiative), MEXT, Japan.
 SH acknowledges support by the DFG cluster of excellence \lq{}Origin and Structure of the Universe\rq{} (\href{http://www.universe-cluster.de}{\texttt{www.universe-cluster.de}}).
 CDF acknowledges support for this work from the National Science Foundation under Grant No. AST-1715611.
 SHS thanks the Max Planck Society for support through the Max Planck Research Group.
 TT acknowledges support by the Packard Foundation through a Packard Research fellowship and by the National Science Foundation through NSF grants AST-1714953 and AST-1906976.
 LVEK is partly supported through an NWO-VICI grant (project number 639.043.308).

\end{acknowledgements}

\section*{Affiliations}
{\small{
$^{1}$ DARK, Niels-Bohr Institute, Lyngbyvej 2, 2100 Copenhagen, Denmark\\
$^{2}$ Physics Department UC Davis, 1 Shields Ave., Davis, CA 95616, USA\\
$^{3}$ STAR Institute, Quartier Agora - All\'ee du six Ao\^ut, 19c B-4000 Li\`ege, Belgium \\
$^{4}$ Exzellenzcluster Universe, Boltzmannstr. 2, D-85748 Garching, Germany \\
$^{5}$ Ludwig-Maximilians-Universit{\"a}t, Universit{\"a}ts-Sternwarte, Scheinerstr. 1, D-81679 M{\"u}nchen, Germany \\
$^{6}$Laboratoire d'Astrophysique, {\'E}cole Politechnique F{\'e}d{\'e}rale de Lausanne (EPFL), Obs. de Sauverny, 1290 Versoix, Switzerland\\
$^{7}$Kavli IPMU (WPI), UTIAS, The University of Tokyo, Kashiwa, Chiba 277-8583, Japan\\
$^{8}$Max-Planck-Institut f{\"u}r Astrophysik, Karl-Schwarzschild-Str. 1, 85748 Garching, Germany \\
$^{9}$Physik-Department, Technische Universit{\"a}t M{\"u}nchen, James-Franck-Stra{\ss}e 1, 85748 Garching, Germany  \\
$^{10}$Academia Sinica Institute of Astronomy and Astrophysics (ASIAA), 11F of ASMAB, No.1, Sect.~4, Roosevelt Rd, Taipei 10617 \\
$^{11}$Department of Physics and Astronomy, University of California, Los Angeles, CA 90095, USA \\
$^{12}$Kavli Institute for Particle Astrophysics and Cosmology and Department of Physics, Stanford University, Stanford, CA 94305, USA \\
$^{13}$Kapteyn Astronomical Institute, University of Groningen, P.O.Box 800, 9700AV Groningen, the Netherlands
}}

\appendix

\section{Planck compressed likelihood}
\label{appendix}
Much of the constraining power of the CMB power spectrum can be compressed in three parameters: the physical density of baryons $\Omega_{\rm b}h^{2}$, which determines relative heights of the peaks in the power spectrum, and two so-called shift parameters that describe two fundamental and directly measured angular scales related to the sound horizon and the Hubble horizon at decoupling. The shift parameters are defined by the following equations:
\begin{eqnarray}
\mathcal{R} & = & \sqrt{\Omega_{\rm m}}\frac{D_{\rm A}(z_{\ast})}{H_{0}^{-1}}\\
\theta_{\ast} & = & \frac{r_{\rm s}(z_{\ast})}{D_{\rm A}(z_{\ast})},
\end{eqnarray}
where $z_{\ast}$ is redshift of decoupling and $D_{\rm A}$ is the comoving angular diameter distance which for flat models considered in this work is given by
\begin{eqnarray}
    D_{\rm A} & =  & \mathrm{c}\int_{0}^{z}\frac{\textrm{d}z}{H(z)}\\
    H^{2}(z)& = & H_{0}^{2}[\Omega_{\rm m}(1+z)^{3}+\Omega_{\rm DE}(z)+\Omega_{\gamma}(1+z)^{4}],
\end{eqnarray}
where $\Omega_{\gamma}$ denotes the density parameter of radiation, i.e. $\Omega_{\gamma}=2.47\times10^{-5}h^{-2}$. 

The comoving sound horizon is given by
\begin{equation}
    r_{\rm s}(z)=\frac{ \mathrm{c}}{\sqrt{3}}\int_{z}^{\infty}\frac{\textrm{d}z}{H(z)\sqrt{1+\frac{3\Omega_{\rm b}}{4\Omega_{\gamma}}(1+z)^{-1}}}.
\end{equation}
Here, an additional contribution to the energy density driving the expansion comes from relativistic neutrinos. The density parameter of relativistic neutrinos $\Omega_{\rm n}$ is given by
\begin{equation}
    \Omega_{\rm n}=N_{\rm eff}\frac{7}{8}\Big(\frac{4}{11}\Big)^{4/3}\Omega_{\gamma},
\end{equation}
where $N_{\rm eff}$ is the effective number of neutrinos with $N_{\rm eff}=3.046$ for the baseline model.

We compute redshift $z_{\ast}$ of decoupling employing the following fitting formula \citep{Hu1996}
\begin{eqnarray}
    z_{\ast} & = & 1047[1+0.00124(\Omega_{\rm b}h^{2})^{-0.738}][1+g_{1}(\Omega_{\rm m}h^{2})^{g_{2}}] \\
    g_{1} & = & 0.0783(\Omega_{\rm b}h^{2})^{-0.238}[1+39.5(\Omega_{\rm b}h^{2})^{0.763}]^{-1}\\
    g_{2} & = & 0.56[1+21.1(\Omega_{\rm b}h^{2})^{1.81}]
\end{eqnarray}

The sound horizon imprinted in galaxy clustering and measured from BAO observations is fixed at the drag epoch when the baryons are released from the Compton drag of the photons. The corresponding drag redshift $z_{\rm d}$ can be calculated using the following fitting function \citep{Hu1996}
\begin{eqnarray}
    z_{\rm d} & = & 1345\frac{(\Omega_{\rm m}h^{2})^{0.251}[1+b_{1}(\Omega_{\rm b}h^{2})^{b_{2}})]}{
     1+0.659(\Omega_{\rm m}h^{2})^{0.828}} \\
     b_{1} & = & 0.313(\Omega_{\rm m}h^{2})^{-0.419}[1+0.607(\Omega_{\rm m}h^{2})^{0.674}]\\
     b_{2} & = & 0.238(\Omega_{\rm m}h^{2})^{0.223}.
\end{eqnarray}

The compressed CMB likelihood is given by a three-dimensional Gaussian distribution in the three parameters mentioned above, i.e. $\Omega_{\rm n}h^{2}$, $\mathcal{R}$ and $\theta_{\ast}$. We employ the mean values and the covariance matrix determined from publicly available MCMC models obtained for a flat $\Lambda$CDM model fitted to the \textit{Planck} observations including the temperature, polarization and lensing data \citep{Planck2018}: $(100\Omega_{\rm b}h^{2},100\theta_{\ast},\mathcal{R})=(2.237\pm0.015,1.0411\pm0.00031,1.74998\pm0.004)$ with the following correlation matrix
\begin{equation}
    \begin{pmatrix} 
   1.00 & 0.34 & -0.63 \\   
  0.34  &      1.00 &    -0.46 \\     
 -0.63   &   -0.46 &      1.00 \\  
\end{pmatrix}
\end{equation}

\begin{figure}
	\centering
	\includegraphics[width=\linewidth]{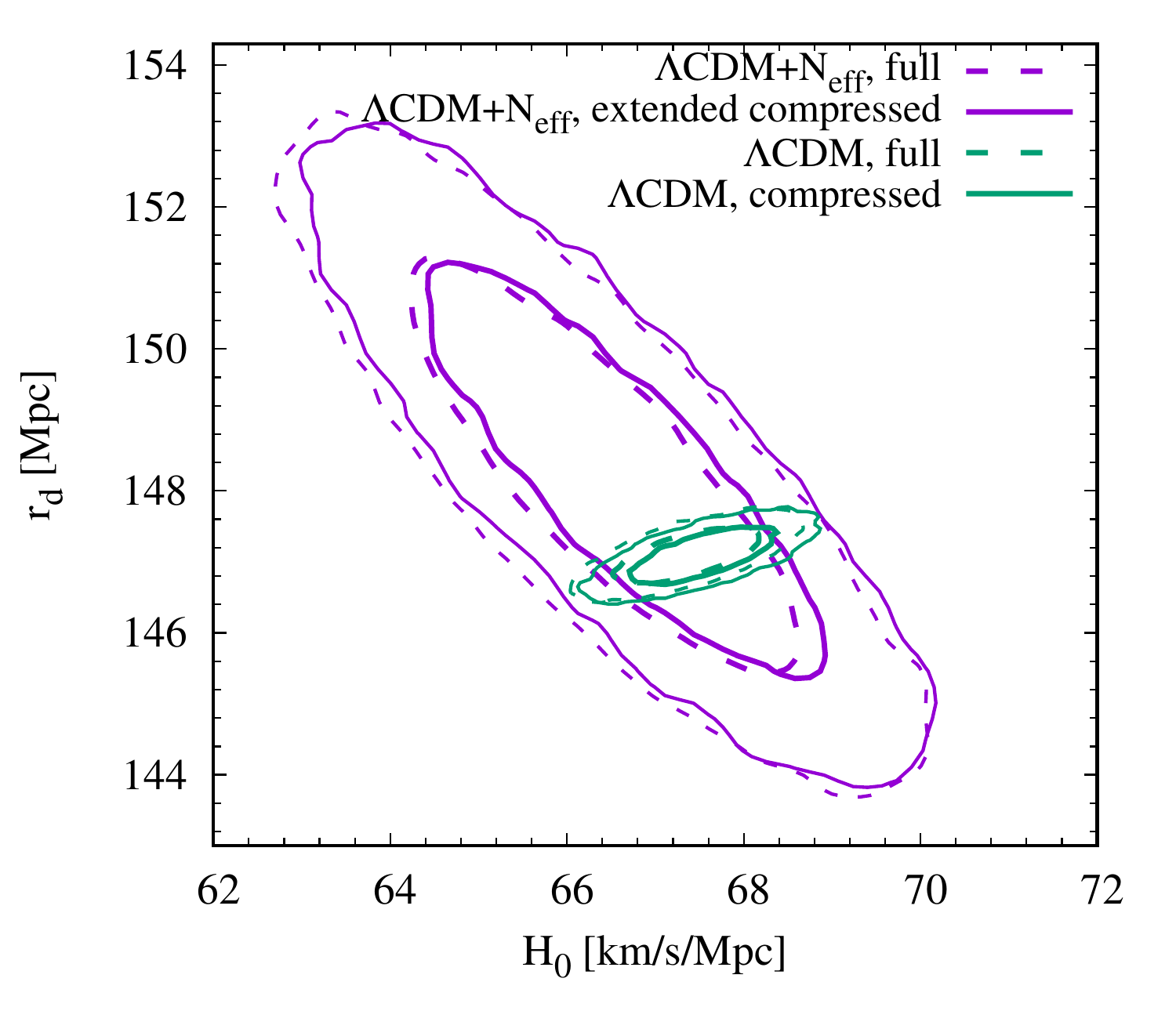}
	\caption{Comparison between constraints on $r_{\rm d}$ and $H_{0}$ from the full \textit{Planck} likelihood (dashed lines) and the compressed likelihood {(for post-recombination modifications of $\Lambda$CDM) or the extended compressed likelihood (for pre-recombination modifications of $\Lambda$CDM)} used in this study (solid lines). The robustness test comprises two cases: the standard flat $\Lambda$CDM model and its extension with a free number of neutrinos.}
	\label{fig:compressed}
\end{figure}

The compressed likelihood recovers accurately the actual constraints obtained from the complete likelihood for a flat $\Lambda$CDM model (see Fig.~\ref{fig:compressed}). Only a fine adjustment of \rev{the redshift scales in both fitting formulae ($\delta z/z\sim 10^{-3}$ smaller relative to the values adopted in \citet{Hu1996})} was applied in order to correct for a sub-percent bias in the mean values of relevant parameters. In general, both approximations used to compute $z_{\ast}$ and $z_{\rm drag}$ are accurate to within 1 per cent in a wide range of the matter and baryon density parameters \citep{Hu1996}.

For early-time extensions of the standard $\Lambda$CDM cosmology (such as a model with free $N_{\rm{eff}}$), the compressed likelihood turns out to be insufficient, leading to a family of models with a wide range of amplitudes of the first peak in the power spectrum. In order to circumvent this problem, we extend the compressed likelihood described above by accounting for the height of the first peak in the power spectrum as an additional constraint. Bearing in mind that the amplitude scales with $\Omega_{\rm dm}h^{2}$, i.e. the physical density of dark matter, a simple extension relies on adding $\Omega_{\rm dm}h^{2}$ as the fourth variable in the compressed likelihood function. Using \textit{Planck} results for a $\Lambda$CDM model with a free effective number of neutrinos as a base early-time extension (inferred from the full temperature and polarization data), we determine the mean values and the covariance matrix of the new four-parameter compressed likelihood, obtaining $(100\Omega_{\rm b}h^{2},100\theta_{\ast},\mathcal{R},\Omega_{\rm dm}h^{2})=(2.225\pm0.0223,1.0414\pm0.00054,1.7529\pm0.0056,0.1184\pm0.0029)$ and the following correlation matrix
\begin{equation}
    \begin{pmatrix} 
   1.00     & -0.50 &      -0.79    &   0.51 \\     
 -0.50 & 1.00 &     0.30 &      -0.81 \\
 -0.79 &     0.30   &     1.00&  -0.19 \\   
  0.51  &    -0.81 &      -0.19 &   1.00     \\
\end{pmatrix}
\end{equation}
Fig.~\ref{fig:compressed} demonstrates that the extended compressed likelihood accurately recovers the actual constraints on $r_{\rm d}$ and $H_{0}$ from \textit{Planck} for a model with a free effective number of neutrinos.

\section{\nikki{Polynomial parametrizations}} \label{Appendix:parametrizations}

\noindent \nikki{This section gives more detailed information about the polynomial parametrizations used throughout this work.}

\nikki{\subsection{Expansion formulas}}

\noindent \nikki{Our first model is the simplest one and adopts a polynomial expansion of $H(z)$ in $z$.
\begin{equation}\label{eq:model1}
{{H(z) = H_0 \left[ 1 + b_1 \, z + b_2 \, z^2 + \mathcal{O}(z^3)\right],}}
\end{equation}
where $H_0$ is the Hubble constant and the coefficient $b_1$ is related to the deceleration parameter $q_0$ through
\begin{equation}
{b_1 = 1 + q_0.}
\end{equation}}
\nikki{In our second model, the luminosity distance $D_L$ is expanded as a polynomial in $\log(1+z)$. \footnote{\nikki{Here, $\log(1+z)$ refers to the log base 10, and not to the natural logarithm.}}}
\begin{align}\label{eq:model2}
&{x = \log(1+z),} \nonumber \\
&{D_L (z) = \frac{c \, \ln(10)}{H_0} \left[ x + c_2 x^2 + c_3 x^3 + c_4 x^4 + \mathcal{O}(x^5) \right],}
\end{align}
\nikki{where the coefficient $c_2$ is related to the deceleration parameter through the following relation:
\begin{equation}
{c_2 = \frac{\ln(10)}{2} \, (2 - q_0).}
\end{equation}
This different parametrization is chosen in order to avoid convergence problems with the Taylor expansion around zero, when employing data with redshifts ${z>1}$. By introducing a new variable ${x}$ that satisfies ${x=0}$ when ${z=0}$ and ${x<1}$ when ${z \rightarrow 2}$ (where the upper limit of 2 is based on the highest lensed quasar redshift), the parametrization is kept within the convergence radius of the Taylor expansion.} \\

\noindent \nikki{Our third model describes transverse comoving distances $D_M$ by polynomials in $z/(1 + z)$. 
\begin{align}\label{eq:model3}
&{y = \frac{z}{1+z},} \nonumber\\
&{D_M (z) = \frac{c}{H_0} \left[ y + d_2 y^2 + d_3 y^3 + d_4 y^4 + \mathcal{O}(y^5) \right],}
\end{align}
where the coefficient $d_2$ is related to the deceleration parameter through
\begin{equation}
{d_2 = \tfrac{1}{2} (1 - q_0).}
\end{equation}
This parametrization is, similar to the one in model 2, chosen to overcome convergence problems.} \\

\nikki{\subsection{Truncation of the polynomials}}

\begin{figure}
	\centering
	\includegraphics[width=\linewidth]{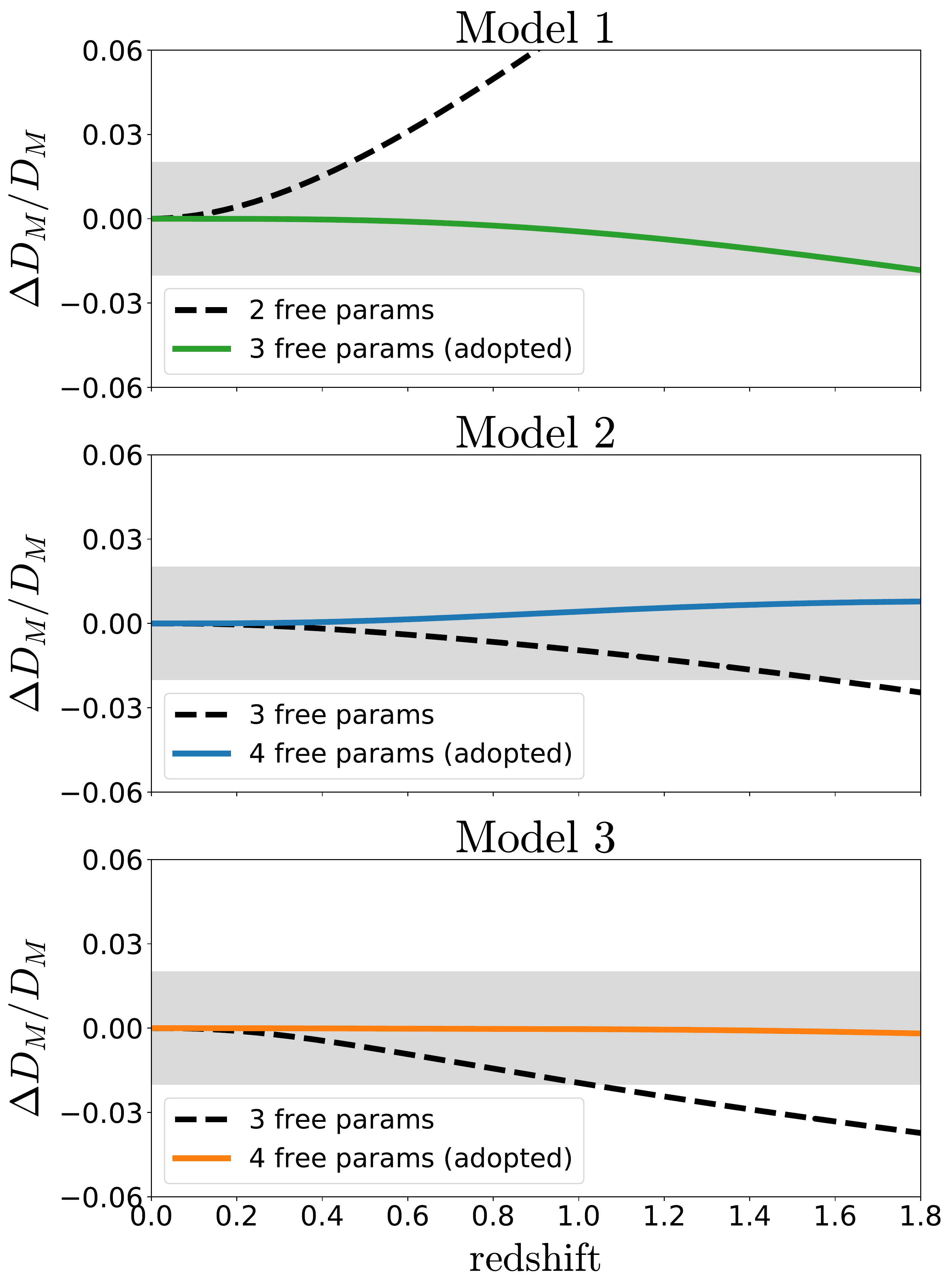}
\caption{{
Relative differences between distances in a fiducial flat $\Lambda$CDM model and distances derived from models 1-3 with free parameters matched to the kinematical coefficients of the fiducial model, $\Delta D_{\rm M} / D_{\rm M} = (D_{\rm{M, expansion}}- D_{\rm{M}, \Lambda\rm{CDM}}) / D_{\rm{M}, \Lambda\rm{CDM}}$. The solid lines show the results satisfying the convergence criterion which sets the truncation of polynomials used in the adopted models in this study.}}
\label{fig:convergence_test} 
\end{figure}

\noindent \nikki{An important thing to consider is at which order the Taylor expansions should be truncated. Higher orders of expansions can give better approximations to the shape of the data, but also introduce more free parameters and therefore larger uncertainties. In order to determine the truncation of the polynomials as given in \ref{eq:model1}, \ref{eq:model2} and \ref{eq:model3}, we perform a convergence test to check that the models can accurately recover expansion history of a fiducial flat $\Lambda$CDM cosmological model in a redshift range of observational data used in our study, i.e. $z<1.8$. The test relies on comparing distances from models 1-3 to the actual distances in the fiducial model. Free parameters of the models are determined by matching coefficients of Taylor expanded Hubble parameter in models 1-3 and the fiducial model. The latter yields well-known kinematical coefficients \citep{Weinberg1972,Visser2004}:}
\begin{align}
&{q_0 = \tfrac{3}{2} \Omega_m - 1,} \nonumber \\
&{j_0 = 1,} \nonumber \\
&{s_0 = 1 - \tfrac{9}{2} \Omega_m.}
\end{align}
\nikki{Since the errors that we obtain by combining calibrations of H0LiCOW and SH0ES are around 2$\%$ (see Table \ref{table2}), we require our models to be within a 2 $\%$ accuracy of $\Lambda$CDM distances in this test. The results can be seen in figure \ref{fig:convergence_test} for $\Omega_{m}=0.3$, where the shaded region corresponds to this imposed limit. It suffices to employ three free parameters (corresponding to a second order polynomial) for model 1 and four free parameters (corresponding to a fourth order polynomial) for models 2 and 3 to satisfy the convergence condition.} {Since a further increase of the number of free parameters is disfavoured by the BIC obtained in fits with the actual late-time observations, these polynomial truncations are adopted in our study (see Table~\ref{table:models}). The BIC score is calculated as
\begin{equation}
{\mathrm{BIC}\ =\ \ln(N)k-2\ln(\mathcal{L}_{m.a.p.}),}
\end{equation}
where $N$ is the number of data points and $k$ is the number of all free parameters in the cosmological fits.} \\

\rev{\subsection{Test with mock distance modulus data}}

\begin{figure}
	\centering
	\includegraphics[width=\linewidth]{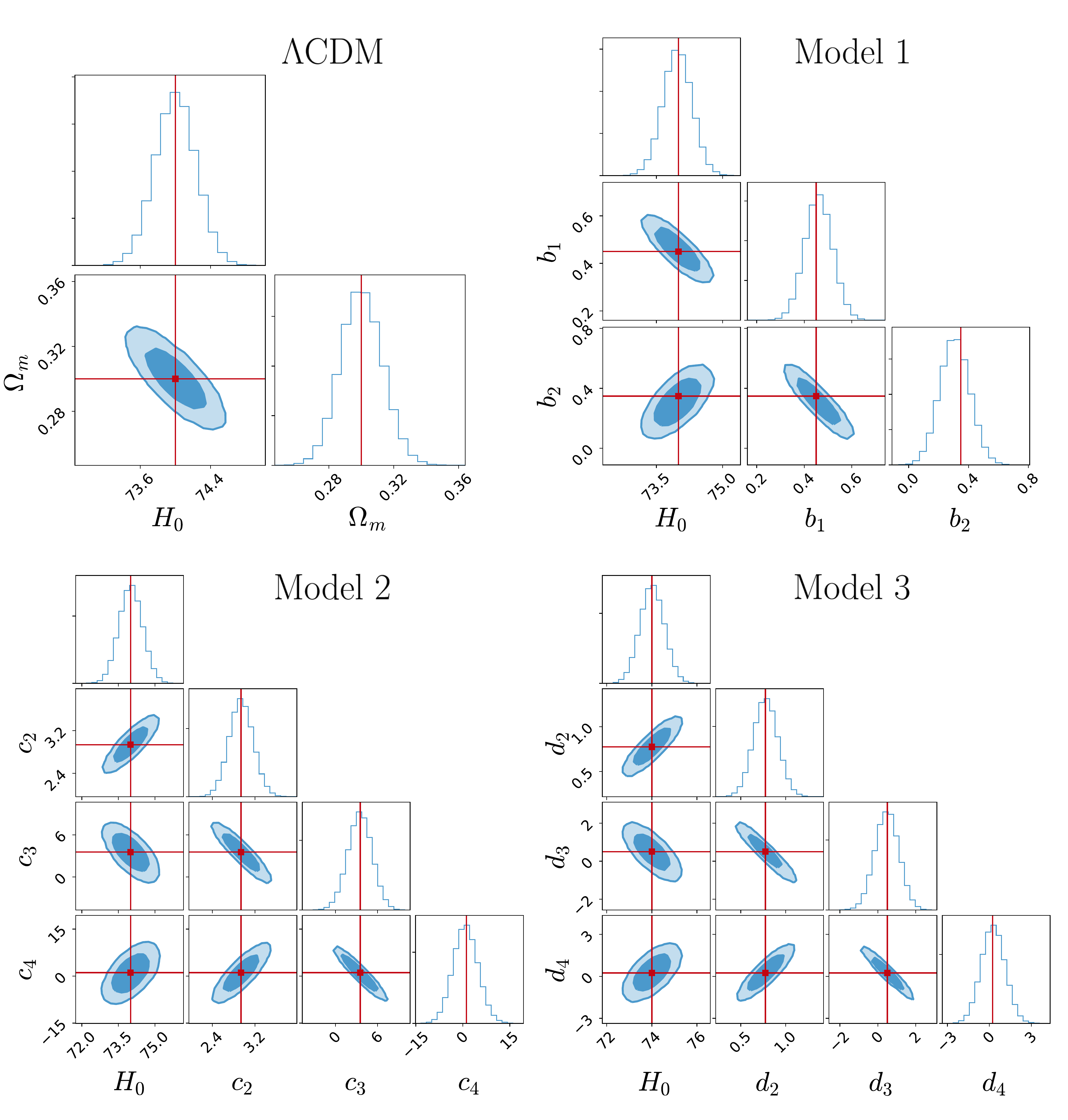}
\caption{\rev{Best-fit values of flat $\Lambda$CDM and polynomial parametrization models 1-3 to mock data. The mock data is generated by replacing the Pantheon distance modulus points by their fiducial flat $\Lambda$CDM values. The red lines indicate the canonical $\Lambda$CDM values of $\Omega_m$, $H_0$ and the expansion coefficients $b_i$, $c_i$ and $d_i$.}}
\label{fig:mocktest} 
\end{figure}

\noindent \rev{As a final test for our polynomial parametrization models, we investigate if any biases are introduced when we fit models 1-3 to flat $\Lambda$CDM data. We transform the Pantheon SN data set to a mock data set, by replacing their binned distance modulus entries by the fiducial flat $\Lambda$CDM values (adopting $H_0 = 74$ km~s$^{-1}$~Mpc$^{-1}$ and $\Omega_m = 0.3$) at the same redshifts. For the errors associated with the distance moduli we keep the original Pantheon ones. By construction, best fit $\Lambda$CDM parameters are equal to their fiducial values, whereas relative shifts in best fit parameters obtained for non-$\Lambda$CDM models measure the corresponding biases.}
\rev{ This test is similar to the one performed by \cite{Yang2019}, in which they find that our model 2 introduces an artificial bias. 
However, their mock data set is based on Pantheon data as well as high-redshift quasar and GRB data (with $z_{\rm{max}} = 6.7$), while in our work we only use sources below $z=1.8$. Figure \ref{fig:mocktest} shows the best-fit values for the coefficients $b_i$, $c_i$ and $d_i$ of models 1-3, obtained with MCMC, and their true values in a flat $\Lambda$CDM cosmology. As can be seen, they are in complete agreement with each other. In fact, the relative difference in $H_0$ between the fiducial value and those of models 1-3 is 0.03\%, 0.02\% and 0.02\%, respectively. This bias is about a hundred times smaller than the current precision achieved by SH0ES and H0LiCOW data (which is around $2\%$). The bias in $q0$ is larger: 2.0\%, 1.2\% and 1.3\% for models 1-3, but still negligible compared to our obtained errors in $q_0$ (which are 10\% at best).}

\rev{This test demonstrates that if the underlying cosmology is flat $\Lambda$CDM, then our models will not introduce any significant biases in the Pantheon redshift range. The convergence test in the previous section also guarantees this. The bias that \cite{Yang2019} found in their model was a consequence of it not passing the convergence test over the complete redshift range of $z=0-7$. }

\rev{We repeat the test for the PEDE model and for a $w$CDM cosmology with $w=-1.2$. In both cases we assume $\Omega_m = 0.3$. We find only a sub-percent bias in the best fit $H_0$ and a few-percent bias in $q_0$ where the actual values are given by}
\begin{align}
&{q_{0, \rm{PEDE}} = \tfrac{3}{2} \Omega_m - \frac{1 - \Omega_m}{2 \ln(10)} - 1,} \nonumber \\
&{q_{0, w\rm{CDM}} = \tfrac{1}{2} + \tfrac{3}{2} w (1 - \Omega_m).}
\end{align}

\onecolumn

\end{document}